\documentclass[reprint,
superscriptaddress,
amsmath,
amssymb,
aps,
prl,
floatfix,
english
]{revtex4-2}

\usepackage{bibunits}
\defaultbibliography{references}
\defaultbibliographystyle{apsrev4-2}
\usepackage{graphicx}
\usepackage{svg}
\usepackage{dcolumn}
\usepackage{bm}
\usepackage{siunitx}
\usepackage[T1]{fontenc}
\usepackage[utf8]{inputenc}

\usepackage[unicode=true, colorlinks=true, citecolor={blue!80!black}, urlcolor={blue!50!black}, linkcolor = {blue!80!black}]{hyperref}

\newcommand{\sctn}[1]{{\textit{#1}}}

\graphicspath{{Figures/}}

\begin{document}
\widetext

\title{Criticality in the crossed Andreev reflection of a quantum Hall edge}

\author{Vladislav D.~Kurilovich}
\email{vlad.kurilovich@yale.edu}
\affiliation{Department of Physics, Yale University, New Haven, CT 06520, USA}
\author{Leonid I.~Glazman}
\affiliation{Department of Physics, Yale University, New Haven, CT 06520, USA}

\begin{abstract}
We develop a theory of the non-local transport of 
two counter-propagating $\nu = 1$ quantum Hall edges coupled via a narrow disordered superconductor.
The system is self-tuned to the critical point between trivial and topological phases by the competition between tunneling processes with or without particle-hole conversion. 
The critical conductance 
is a random, sample-specific quantity with a zero average and unusual bias dependence.
The negative values of conductance are relatively stable against variations of the carrier density, which may make the critical state to appear as a topological one. 
\end{abstract}

\maketitle

\sctn{Introduction.}---Topological superconductivity provides a promising route to the fault-tolerant quantum computing \cite{nayak2008}. 
A one-dimensional topological superconductor hosts non-Abelian excitations at its ends, such as Majorana zero modes \cite{kitaev2001} or their fractional generalizations, i.e., parafermions \cite{clarke2013, meng2012}.
The non-local character of these modes can be harnessed to store quantum information in a way inherently protected from the decoherence.
One can manipulate the protected information by braiding the zero modes \cite{ivanov2001}, thanks to their non-Abelian exchange statistics.

There are many proposed implementations of a one-dimensional topological superconductor (see, e.g., \cite{lutchyn2010, oreg2010, nadj-perge2013, pientka2017}).
A versatile platform that may 
host Majorana zero modes (or parafermions) is a hybrid quantum Hall-superconductor structure \cite{clarke2013, clarke2014, mong2014}.
The basic idea behind it is to couple two counter-propagating quantum Hall edges via a conventional superconductor. 
If the width of the superconductor $d$ is comparable to the coherence length $\xi$, then two electrons residing in different edges can transfer into the superconductor as a Cooper pair.
This process can be viewed as a non-local counterpart of the Andreev reflection, and is thus called a crossed Andreev reflection (CAR).
CARs establish superconducting correlations between the edges.
At filling $\nu = 1$, which we shall focus on, the induced pairing has the $p$-wave symmetry, as required for the topological superconductivity supporting Majorana zero modes.
In fact, experiments on quantum transport in such setups~\cite{lee2017, gul2021} (as well as in the related ones~\cite{zhao2020, hatefipour2022}) have already started. This motivated a lot of recent theoretical works \cite{manesco21, kurilovich2022, schiller2022, galambos2022, michelsen2022, tang2022}.

A complication in reaching the topological phase arises due to the elastic cotunneling (EC) processes which compete with CAR~\cite{beckmann2004, cadden-zimansky2006}.
In an EC event, a particle tunnels across the superconductor without a conversion to a hole (contrary to the CAR event). This amounts to an electron backscattering process.
A strong backscattering is detrimental for the topological phase~\cite{brouwer2011}.

The competition between
CAR and EC processes is sensitive to disorder. 
We note that only a superconductor with a high upper critical field $H_{\rm c2}$ is compatible with the quantum Hall effect.
This dictates the use of ``dirty'' superconductors with the electron mean free path $l_{\rm mfp}\ll \xi$
\cite{lee2017, gul2021}.
For a dirty superconductor, EC and CAR have, at best, the same probabilities \cite{falci2001, bignon2004}.
Due to the spin-polarization of the $\nu = 1$ state, the particular relation between the two processes is determined by the strength of the spin-orbit interaction in the system.
If the spin-orbit interaction is weak, then CAR processes are largely inhibited and EC prevails.
This brings the proximitized edge states into a trivial phase.
A sufficiently strong spin-orbit interaction allows the probabilities of CAR and EC to approach each other. 
As we show, this naturally tunes the system to the critical point of the transition between trivial and topological phases.

\begin{figure}[t]
  \begin{center}
    \includegraphics[scale = 1]{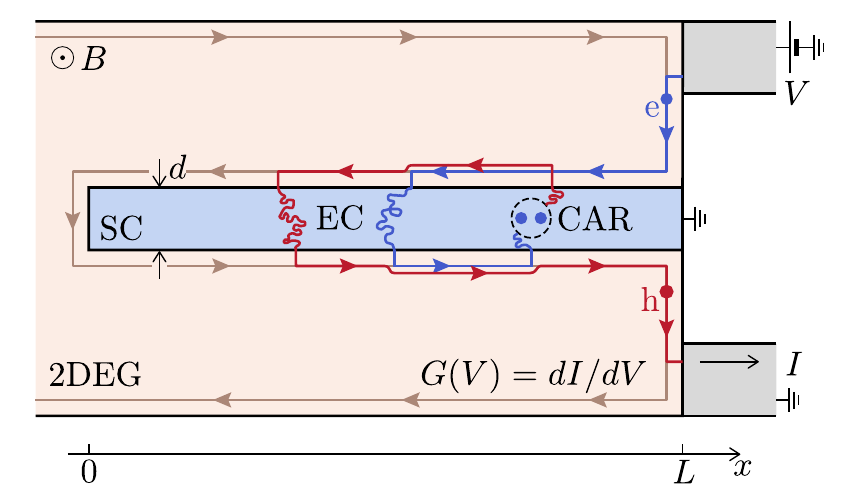}
    \caption{Schematic layout of the considered setup.
    A narrow superconductor induces the proximity effect into two counter-propagating chiral edge states.
    Electrons are incident onto the superconductor from an upstream electrode biased by voltage $V$.
    Non-local conductance $G(V)$ is determined by the interference of elastic cotunneling and crossed Andreev reflection processes.
    These processes are random but are balanced statistically [see Eqs.~\eqref{eq:CAR_EC_short}--\eqref{eq:crit_condition}], which sets the proximitized edge states at the critical point between trivial and topological phases.
    }
    \label{fig:setup}
  \end{center}
\end{figure}

The critical point belongs to the infinite-randomness universality class \cite{motrunich2001}.
Although many authors studied the thermal transport of a superconductor in a critical state \cite{brouwer2000, akhmerov2011, antonenko2020}, the charge transport has attracted surprisingly little attention (with a notable exception of Ref.~\onlinecite{shivamoggi2010}). 
The conductance $G = dI / dV$ associated with the backscattered current $I$ (see Fig.~\ref{fig:setup}) is random.
Its dependence on bias $V$ is not well-understood even at the smallest biases.
Besides, little is known about the variations of the conductance with the device parameters such as the electron density.
Such parametric dependence is of direct relevance to the experiments, as the electron density is one of the simplest knobs which tunes the properties of the quantum Hall state.
In this work, we address the bias- and density-dependence of the proximitized edge states conductance.

At the critical point, the dependence of conductance on bias is a stochastic function alternating between positive and negative values.
The pattern of these fluctuations is determined by a realization of the disorder in the superconductor.
The criticality is reflected in the unusual logarithmic scaling of the differential conductance correlation function, which we quantify.

At any bias, the ensemble-averaged conductance is zero. 
The ensemble averaging can be achieved in a given device by varying its parameters, such as the electron density \cite{lee1987}.
Interestingly, we find that the variation of the CAR amplitude happens at a much larger density scale than that of the EC amplitude.
If---due to a statistical fluctuation---the former amplitude is relatively large, then the conductance may stay negative in a broad range of densities.
This conclusion may help in interpreting recent experiments \cite{lee2017, gul2021}, in which a negative conductance stable to the variation of density was observed.
Our theory shows that an observation of a negative conductance does not imply a topological phase by itself; rather, it may be a facet of a critical state of proximitized edges.

\sctn{Model.}---We consider two counter-propagating $\nu~=~1$ quantum Hall edge states coupled through a narrow superconducting electrode, see Fig.~\ref{fig:setup}. At low temperatures and bias, transport of the edge states can be described with the help of an effective low-energy Hamiltonian: $H_{\rm eff} = H_{\rm QH} + H_{\rm prox}$. 
The first term is the Hamiltonian of electron modes propagating along each of the edges:
\begin{equation}\label{eq:H_QH}
    H_{\rm QH} = \sum_{j = R,L}\int dx\,\psi^\dagger_j(x) v[-i\sigma_j \partial_x - k_\mu] \psi_j(x),
\end{equation}
where $\psi_{j}(x)$ is the field operator of a right ($j = R$) or left ($j = L$) moving chiral electron, $\sigma_{R/L} = \pm 1$, $v$ is the edge states Fermi velocity, and $k_\mu$ is their Fermi momentum.

The term $H_{\rm prox}$ describes the coupling between the edge states through the superconductor.
We find it in the same way as in Ref.~\onlinecite{kurilovich2022}: we start with the tunneling Hamiltonian $H_{\rm T}$ for the coupling of each of the edge states
with the superconductor, and then ``integrate out'' the superconductor's degrees of freedom. For electron energies $E \ll \Delta$ measured with respect to the Fermi level, the procedure results in
\begin{align}\label{eq:H_prox}
    H_{\rm prox}\!=\!\sum_{i,j=R,L} \frac{t^2}
    {2}\int\! dxdx'\,
    \hat{\psi}_i^\dagger(x)
    \partial^2_{y_i y_j}
    {\cal G}(x,y_i;x^\prime,y_j)\,
    \hat{\psi}_j(x').
\end{align}
Here $\hat{\psi}_i(x) = \bigl(\psi_i(x), \psi_i^\dagger(x)\bigr)^T$, $y_R = 0$, $y_L = d$, and $d$ is the electrode width. Properties of the superconductor---such as its energy gap $\Delta$ and spin-orbit coupling in it---are encoded into the superconductor Green's function ${\cal G}$, which is a $2 \times 2$ matrix in the Nambu space.
The normal derivatives are computed at the respective interfaces $y = 0, d$  \cite{prada2004, lutchyn2012}. $t$ is the tunneling amplitude between each of the edge states and the superconductor.

We are interested in the differential conductance $G(V)$ of the three-terminal setup depicted in Fig.~\ref{fig:setup}. A chiral electron incident on the superconducting electrode can be transmitted across it either as a particle or---if a crossed Andreev reflection happens---as a hole. By labelling the amplitudes of these processes at energy $E$ as $A_{\rm N}(E)$ and $A_{\rm A}(E)$, respectively, we can express $G(V)$ at zero temperature as
\begin{equation}\label{eq:conductance}
    G(V, T = 0) = G_Q \bigl(|A_{\rm N}(E)|^2 - |A_{\rm A}(E)|^2\bigr)\bigr|_{E = eV},
\end{equation}
where $G_Q = e^2 / 2 \pi$ is the conductance quantum (we use units with $\hbar = 1$). One can view the two needed amplitudes as the entries of a scattering matrix $S(E)$, relating the incoming and outgoing electron and hole waves: 
$A_{\rm N / A}(E) \equiv S_{\rm ee / he}(E)$. Therefore, we need to find $S(E)$ to determine the conductance. A convenient way to do that is to first divide the electrode into a sequence of short elements (short enough to be treated perturbatively), and then track how $S(E)$ changes as we ``build'' the electrode by stacking the elements together.

\sctn{Scattering off a short element.}---A single element acts as a bridge between the two chiral edges. 
As an electron traverses the bridge, it either turns into a hole (a CAR process), or remains an electron (an EC process).
For a short element, we find the amplitudes of the two processes treating $H_{\rm prox}$ of Eq.~\eqref{eq:H_prox} in the Born approximation. We obtain for CAR and EC amplitudes at the Fermi level, respectively:
\begin{subequations}\label{eq:CAR_EC}
\begin{align}
    \delta A_{\rm A}\hspace{-0.1cm} &=\hspace{-0.07cm} \frac{t^2}{v}\hspace{-0.1cm}\int\hspace{-0.1cm}  dxdx^\prime e^{-ik_\mu(x^\prime - x)} \partial^2_{yy^\prime}{\cal G}_{\rm he}^{\downarrow\uparrow}(x, 0;x^\prime, d),\label{eq:CAR_basic}\\
    \delta A_{\rm N}\hspace{-0.1cm} &= \hspace{-0.07cm} \frac{t^2}{v}\hspace{-0.1cm} \int\hspace{-0.1cm}  dxdx^\prime e^{-ik_\mu(x^\prime + x)} \partial^2_{yy^\prime}{\cal G}_{\rm ee}^{\uparrow\uparrow}(x, 0;x^\prime, d).\label{eq:EC_basic}
\end{align}
\end{subequations}
Here ${\cal G}_{\rm he}$ and ${\cal G}_{\rm ee}$ are the anomalous and normal components of the Green's function of the superconductor \cite{landau-9}. The spin of a hole in a $\nu = 1$ edge is opposite to that of an electron. Thus, for a CAR to happen the quasiparticle has to flip its spin upon traversing the element, as indicated by the $\uparrow \downarrow$ superscript in Eq.~\eqref{eq:CAR_basic}. In our model, the spin-flip processes result from the spin-orbit scattering in the superconductor. 

The Green's functions in Eqs.~\eqref{eq:CAR_basic} and \eqref{eq:EC_basic} describe the propagation of an electron wave across the superconducting element. Due to the diffraction of the wave on the impurities in the superconductor, the result of the wave propagation is stochastic. 
Therefore, ${\cal G}_{\rm ee / he}$ and $\delta A_{\rm A/ N}$ are random quantities.
We shall characterize their properties in an experimentally relevant regime of the electron mean free path $l_{\rm mfp} \ll d$ \cite{lee2017, gul2021}. Under the latter condition, we can estimate $\langle {\cal G}_{\rm ee / he} \rangle \propto \exp(- d / 2l_{\rm mfp}) \ll 1$, where $\langle \dots \rangle$ denotes the average over the disorder configurations in the superconductor. 
In what follows, we neglect these exponentially small quantities and approximate $\langle \delta A_{\rm A} \rangle = \langle \delta A_{\rm N} \rangle = 0$.

Next, we find the variances $\langle |\delta A_{\rm A / N}|^2 \rangle$ of the CAR and EC amplitudes. This requires averaging the product of the two Green's functions of the superconductor. Such an averaging can be performed by relating $\langle {\cal G} \cdot {\cal G}\rangle$ to the normal-state diffuson via a standard procedure~\cite{hekking1994}. 
Focusing on the ``dirty'' superconductor and taking the spin-orbit scattering into the account \cite{abrikosov1962, vavilov2003}, we find for an element of length $\delta L \gg \xi$ and width $d \gg \xi$:
\begin{equation}\label{eq:CAR_EC_short}
\langle |\delta A_{\rm A / N}|^2 \rangle = \frac{\delta L}{l_{\rm A / N}}.
\end{equation}
The length scales $l_{\rm A}$ and $l_{\rm N}$ are given by \cite{sm}
\begin{equation}\label{eq:CAR_EC_final}
    \frac{1}{l_{\rm A / N}} = \frac{4\pi g^2}{G_Q\sigma} \sqrt{\frac{\pi \xi}{2 d}} \left(e^{-d / \xi} \mp e^{-d / \xi^\star}\sqrt{{\xi^\star/\xi}}\right).
\end{equation}
Here $g = 2\pi^2 G_Q p_F \nu_{\rm M} \nu_{\rm edge} t^2$ is the conductance per unit length of the interface between the chiral edge and the electrode in the normal state. The conductance depends on the Fermi momentum of the metal $p_F$ and the density of states $\nu_{\rm M}$ in it, in addition to its dependence on the tunneling amplitude $t$ and the density of states $\nu_{\rm edge} = 1/(2\pi v)$ at the edge. The dependence of $l_{\rm A / N}$ on the normal state conductivity of the metal $\sigma$ reflects the diffusive character of electron motion in the dirty superconductor. Finally, $\xi^\star = \xi / \sqrt{1 + 4 / (3 \tau_{\rm so} \Delta)}$, where $\tau_{\rm so}$ is the spin-orbit scattering time.
Since a spin-flip is needed for a CAR of a spin-polarized edge, $\delta A_{\rm A} = 0$ in the absence of spin-orbit scattering. This is why $1/l_{\rm A} = 0$ if $\xi = \xi^\star$.

Equations \eqref{eq:CAR_EC_short} and \eqref{eq:CAR_EC_final} show that, generically, $\langle |\delta A_{\rm A}|^2 \rangle \leq \langle |\delta A_{\rm N}|^2\rangle$.
The limit of strong spin-orbit coupling, $\tau_{\rm so} \Delta \ll 1$, is the most favorable one for the topological superconductivity.
In this limit, the electron spin fully randomizes in the course of tunneling, which results in $\langle |\delta A_{\rm A}|^2 \rangle = \langle |\delta A_{\rm N}|^2\rangle$. Equivalently,
\begin{equation}\label{eq:crit_condition}
l_{\rm A} = l_{\rm N} = 2l_0
\end{equation} (we introduced the factor of two for the notational convenience).
We will demonstrate below that, in fact, the latter condition is \textit{insufficient} to reach the topological phase; instead, it corresponds to a critical point of the transition between trivial and topological phases. 

In addition to the non-local CAR and EC processes, an electron can return to the same edge after its excursion in the superconductor. This leads to accumulation of the forward scattering phase $\delta \Theta_j$ ($j = R, L$). We find $\langle \delta \Theta_j\rangle = 0$ and $\langle \delta \Theta_j^2 \rangle = \delta L / l_{\rm F}$. The particular expression for $l_{\rm F}$ is inconsequential for our conclusions and we relegate it to the supplement~\cite{sm}. We note that the Pauli exclusion principle forbids an Andreev reflection within a single $\nu = 1$ edge at the Fermi level. This is in contrast to the case of $\nu = 2$ considered in Ref.~\onlinecite{kurilovich2022}.

Finally, only a type II superconductor is compatible with the magnetic field required for reaching the quantum Hall state.
The field may induce vortices, whose normal cores give rise to the non-vanishing density of states at the Fermi level in the superconductor~\footnote{Vortices are induced in the superconducting electrode if the magnetic field exceeds the width-dependent lower critical field $H_{\rm c 1}(d)$. For a narrow electrode, $d \sim \xi$, the latter field is comparable to the upper critical one~\cite{abrikosov1964, likharev1971}.}.
As a result, a quasiparticle incident on the superconducting element can tunnel into it normally, irreversibly leaving the edges. 
We model such a quasiparticle loss phenomenologically by assigning to each element a loss probability $\delta p = 4 \Gamma \delta L / v$ proportional to its length. Parameter $\Gamma$ has a meaning of the rate at which the edge quasiparticles are lost.

\sctn{Scattering matrix of a long electrode.}---An electron incident on a long ($L \gtrsim l_0$) superconducting electrode undergoes multiple CAR and EC processes. In this case, one cannot directly apply the perturbation theory to find the scattering matrix. Instead, we break the electrode in a series of short elements labelled by their $x$-coordinate, and track how $S(E)$ evolves as we join the elements together. Equation \eqref{eq:CAR_EC_short} suggests to parameterize the scattering amplitudes of individual elements as
$\delta A_{\rm A / N}(x) = \eta_{\rm A / N}(x) \cdot \sqrt{\delta L} / 2$; 
similarly, we represent $\delta \Theta_j = \vartheta_j(x)\cdot\sqrt{\delta L}$. Random variables $\eta_{m}(x)$ ($m \in \{\rm A, N \}$) and $\vartheta_j(x)$ ($j \in \{\rm R, L \}$)
are Gaussian, mutually independent, and uncorrelated for different $x$. Using Eq.~\eqref{eq:CAR_EC_short} and a respective relation for the forward scattering phase, 
we find for their correlators:
\begin{equation}\label{eq:correlators}
   \langle \eta_{m}(x) \eta_{m^\prime}^\star(x^\prime) \rangle = \frac{4\delta_{mm^\prime}}{l_{m}} \delta(x - x^\prime)
\end{equation}
and $\langle \vartheta_j(x) \vartheta_{j^\prime}(x^\prime) \rangle = \delta_{jj^\prime} \delta(x - x^\prime) / l_{\rm F}$ \footnote{We promoted the variables $\eta_{m, \alpha}(x)$ and $\vartheta_j(x)$ defined on a discrete set of elements to the continuous fields. This is why we use $\delta(x-x^\prime)$ instead of $\delta_{xx^\prime}$.}.

By evaluating the change of the scattering matrix upon an addition of a single short element to the electrode, we obtain the following equation for the evolution of $S(E)$ with $x$ \cite{sm}:
\begin{equation}\label{eq:langevin_S}
    i\frac{dS}{dx} =-\frac{2(E+i\Gamma)}{v}S+{\cal U} + S\,{\cal U}^\dagger S + S{\cal L} + {\cal R}S.
\end{equation}
Here
\begin{equation}
{\cal U}(x)=\frac{i}{2}\bigl[\eta_{{\rm N}y}(x) \tau_0 - i\eta_{{\rm N}x}(x)\tau_z  + \eta_{{\rm A}x}(x) \tau_y + \eta_{{\rm A}y}(x) \tau_x],
\end{equation}
in which we introduced $\eta_{m x} \equiv {\rm Re}\,\eta_m$ and $\eta_{m y} \equiv {\rm Im}\,\eta_m$; $\tau_{x,y,z}$ are the Pauli matrices in the Nambu space, and $\tau_0$ is the identity matrix. ${\cal R}(x) = -\tau_z \vartheta_R(x)$ and ${\cal L}(x) = -\tau_z \vartheta_L(x)$. The initial condition is $S(E, x = 0) = \tau_0$. 
A combination of terms ${\cal U}$ and $S{\cal U}^\dagger S$ describes the interference between the two paths in which an electron can tunnel from a right-moving edge into a left-moving one, see Fig.~\ref{fig:setup}. The interference leads to the localization of wave functions of the edges.
Note that the quasiparticle loss $\Gamma$ plays the role of the level broadening. 

\sctn{Criticality.}---
We now apply Eq.~\eqref{eq:langevin_S} at $\Gamma = 0$ to reveal the critical behavior of the proximitized edges. 
We demonstrate the criticality by finding the low-energy density of states (DOS) $\nu (E)$ and conductance $G(E)$ in the limit of infinite length $L$ \footnote{We use $G(E)$ as a short-hand notation for the differential conductance $G(V, T = 0)$ at bias $V = E / e$.}.
To allow for small deviations from the critical point, we modify Eq.~\eqref{eq:crit_condition} by taking
$l_{\rm A} = 2l_0 (1 + \lambda)$, $l_{\rm N} = 2l_0 (1 - \lambda)$ with $|\lambda| \ll 1$.

A particularly convenient parameterization of the $S$-matrix for analyzing Eq.~\eqref{eq:langevin_S} is
\begin{subequations}\label{eq:params}
\begin{align}
    S &= 
    \frac{1}{2}
    \begin{pmatrix}
    F_+(w_1,w_2) e^{i\alpha} & F_-(w_1,w_2)e^{i\phi}\\
    F_-(w_1,w_2)e^{-i\phi} & F_+(w_1,w_2)e^{-i\alpha}
    \end{pmatrix},\\
    &F_\pm(w_1, w_2) = -\tanh w_1 + \frac{i}{\cosh w_1} \notag \\
    &\quad\quad\pm{\rm sign}(w_1 - w_2) \Bigl[-\tanh w_2 + \frac{i}{\cosh w_2} \Bigr].\label{eq:Apm}
\end{align}
\end{subequations}
Variables $w_{1,2}$ here are defined in the interval $(-\infty, +\infty)$. Using this parameterization in Eq.~\eqref{eq:langevin_S}, we obtain a system of equations governing the evolution of $w_1, w_2, \alpha$, and~$\phi$:
\begin{subequations}\label{eq:langevin_big}
\begin{align}
    \hspace{-0.3cm}\frac{dw_1}{dx} &= \frac{2E}{v}\cosh w_1 + \eta_{{\rm N} x} \sin \alpha - \eta_{{\rm N} y} \cos \alpha\notag\\
   \hspace{-0.3cm} &\hspace{3.0cm}+ \eta_{{\rm A} x} \sin \phi - \eta_{{\rm A} y} \cos \phi, \label{eq:u_evol}
    \\ 
    \hspace{-0.3cm}\frac{dw_2}{dx} &= \frac{2E}{v}\cosh w_2 + \bigl(\eta_{{\rm N} x} \sin \alpha - \eta_{{\rm N} y} \cos \alpha\notag\\
   \hspace{-0.3cm} &\hspace{1.0cm}- \eta_{{\rm A} x} \sin \phi + \eta_{{\rm A} y} \cos \phi\bigr)\,{\rm sign}(w_1 - w_2), \label{eq:v_evol}\\
   \hspace{-0.3cm} \frac{d\alpha}{dx} &= \vartheta_{\rm R} + \vartheta_{\rm L}\hspace{-0.05cm} + q(w_1, w_2) \bigl(\eta_{{\rm N}x} \cos \alpha + \eta_{{\rm N}y} \sin \alpha\bigr), \label{eq:alpha_evol}\\
    \hspace{-0.3cm}\frac{d\phi}{dx} &= \vartheta_{\rm R} - \vartheta_{\rm L} \hspace{-0.05cm} - \hspace{-0.05cm} q(w_2, w_1)\, {\rm sign}(w_1 - w_2) \notag\\
    &\hspace{3cm}\times\bigl(\eta_{{\rm A}x} \cos \phi + \eta_{{\rm A}y} \sin \phi\bigr). \label{eq:phi_evol}
\end{align}
\end{subequations}
Here we abbreviated $q(w_1, w_2) = \tanh \tfrac{w_1 + w_2}{2}\, \Theta(w_1 - w_2) + \coth \tfrac{w_1 - w_2}{2}\, \Theta(w_2 - w_1)$, with $\Theta(z)$ being the Heaviside step function. The initial conditions for Eqs.~\eqref{eq:u_evol}--\eqref{eq:phi_evol} are $w_1(x = 0) = -\infty$, $w_2(0) - w_1(0) = \varepsilon$ (offset $\varepsilon > 0$ is needed to remove the ambiguity of ${\rm sign}(w_1 - w_2)$ in Eq.~\eqref{eq:Apm}),  and $\alpha (0) = \phi(0) = 0$.

\begin{figure}[t]
  \begin{center}
    \includegraphics[scale = 1]{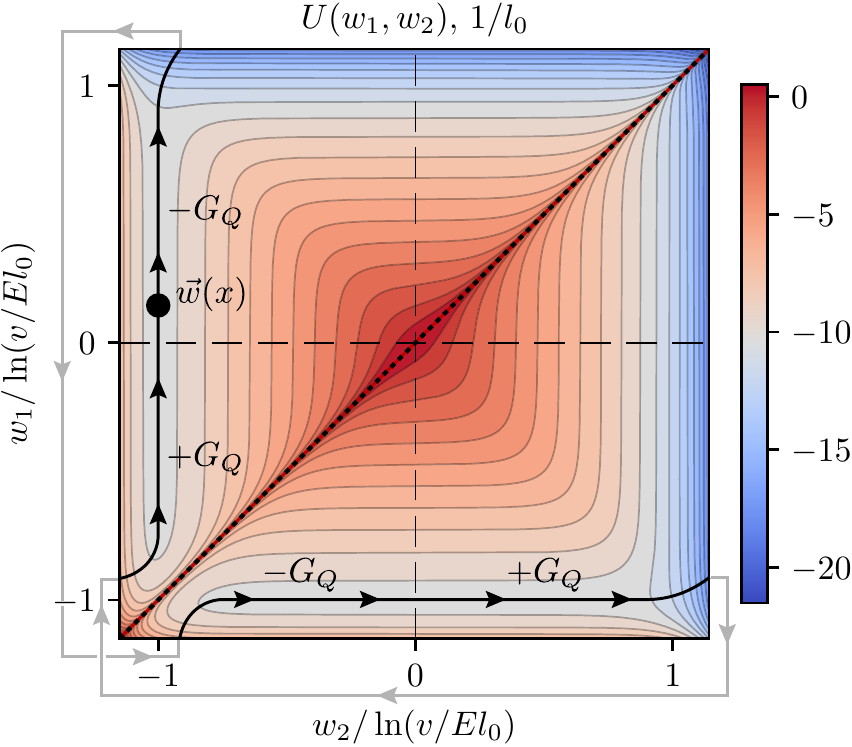}
    \caption{
    Effective potential $U(w_1, w_2)$ for variables $w_1$, $w_2$ which parameterize the scattering matrix $S$ [see Eq.~\eqref{eq:potential}]. The potential landscape is plotted for
    $\ln (v / E l_0) = 12$ at the critical point, $\lambda = 0$. The motion of the $\vec{w}$-particle is cyclic (connected black and grey lines). It is confined to two nearly equipotential trenches: a vertical one at $w_2 = -\ln (v / E l_0)$ and a horizontal one at $w_1 = -\ln (v / E l_0)$. 
    The low-energy conductance takes quantized values $\pm G_Q$ depending on the position of $\vec{w}(L)$ in the $(w_1, w_2)$-plane.}
    \label{fig:potential}
  \end{center}
\end{figure}

We will see momentarily that out of four variables only  $w_1$ and $w_2$ are important for finding 
$\nu(E)$ and 
$G(E)$.
This makes it convenient to derive a simplified system of equations that would focus exclusively on variables $w_1$ and $w_2$.
To do that, we first solve Eqs.~\eqref{eq:alpha_evol} and \eqref{eq:phi_evol} in a short interval $dx$, and then substitute the solutions into Eqs.~\eqref{eq:u_evol} and \eqref{eq:v_evol}.
This leads to \cite{sm}
\begin{align}\label{eq:langevin}
    \frac{dw_i}{dx} = - \frac{\partial U(w_1,w_2)}{\partial w_i} + \tilde{\eta}_i(x).
\end{align}
This equation is similar to the Langevin equation for a Brownian particle moving in an external field.
The noise terms $\tilde{\eta}_i(x)$ are stemming from $\eta_{\rm N}(x)$ and $\eta_{\rm A}(x)$ of Eq.~\eqref{eq:langevin_big}; their correlators are $\langle\tilde{\eta}_i (x) \tilde{\eta}_j(x^\prime)\rangle = \frac{1}{l_0}\delta_{ij}\delta(x - x^\prime)$. 
The first term in Eq.~\eqref{eq:langevin} describes the drift of the $\vec{w}$\,-``particle'' in an external field. The potential of the field is given by
\begin{align}\label{eq:potential}
    U(w_1, &w_2) = -\frac{2E}{v} (\sinh w_1 + \sinh w_2)
    \notag\\
   &- \frac{1}{l_0} \ln |\sinh w_1 - \sinh w_2| + \frac{\lambda}{l_0}(w_1 - w_2).
\end{align}
The potential is plotted in Fig.~\ref{fig:potential}. In the low-energy limit, $E \ll v/ l_0$, and for $|\lambda| \ll 1$, the dynamics of the $\vec{w}$-particle is confined to two elongated trenches of length $2\ln(v / E l_0)\gg 1$. Both trenches end with a ``cliff'': the potential rapidly and boundlessly drops down for $w_{1,2} \gtrsim \ln(v / E l_0)$ due to the first term in Eq.~\eqref{eq:potential}. 
The motion of the $\vec{w}$-particle is cyclical. In one part of the cycle, $\vec{w}$ moves along the vertical trench until it reaches the cliff. 
When it falls off the cliff, it reemerges on the opposite side of the $(w_1, w_2)$-plane, see Fig.~\ref{fig:potential}.
After that a complementary part of the cycle starts in which variables $w_1$ and $w_2$ trade places. 
The jump in one variable occurs at a fixed value of another variable, as required by the continuity of the $S$-matrix as a function of $x$.

The conductance $G(E)$ is determined by the end-point of the $\vec{w}$-particle's evolution. Using Eq.~\eqref{eq:conductance} and Eq.~\eqref{eq:params}, we can express $G(E)$ in terms of the values of $w_1$ and $w_2$ at $x = L$ as
\begin{equation}\label{eq:cond}
    G(E) = G_Q\,{\rm sign}(w_1(L) - w_2(L))\tanh w_1(L) \tanh w_2(L).
\end{equation}
In the low-energy limit, one can approximate $\tanh w_i(L)$ by ${\rm sign}\,w_i(L)$. Indeed, the two functions differ from each other only in a narrow interval of $w_i$ near $w_i = 0$.
The width of this interval $\sim 1$ is much smaller than the lengths of the trenches $\sim \ln (v / E l_0) \gg 1$ for the motion of the $\vec{w}$-particle.
This shows that the low-energy conductance is quantized, $G(E) = \pm G_Q$; the accuracy of quantization is controlled by a small parameter $1 / \ln (v / E l_0) \ll 1$. 
The dependence of the quantized value of $G(E)$ on $\vec{w}(L)$ is illustrated in Fig.~\ref{fig:potential}.

The integrated density of states $N(E) = \int_0^E dE^\prime \nu(E^\prime)$ can also be related to the $S$-matrix \cite{brouwer2011-distr}:
\begin{equation}\label{eq:N_E}
    N(E) = \frac{1}{L}\frac{1}{2\pi i} \ln \det S(E, L).
\end{equation}
Due to the unitarity of the scattering matrix, $\det S = e^{i\beta}$. 
Equation~\eqref{eq:params} shows that the phase $\beta$ winds by $2\pi$ every time the particle completes a full cycle in the $(w_1, w_2)$-plane. Consequently, to find $N(E)$ we need to determine the number of cycles made by the particle per unit ``time''~$x$.

The potential $U$ changes linearly along the trenches due to the third term in Eq.~\eqref{eq:potential}.
If $\lambda > 0$, then the particle diffuses uphill along the vertical trench and downhill along the horizontal one.
As a result, it spends the majority of time being trapped in the potential well near the point $(- \ln (v/El_0), - \ln (v/El_0))$ on the vertical trench.
In the limit $E \rightarrow 0$, the potential well becomes infinitely deep and, according to Eq.~\eqref{eq:cond}, the conductance distribution function approaches $P(G) = \delta(G - G_Q)$.

The configuration reverses for $\lambda < 0$ \footnote{We remind that only the condition $\lambda > 0$ arises naturally for a disordered superconductor, see the discussion around Eq.~\eqref{eq:crit_condition}. It is nonetheless instructive to consider the case $\lambda < 0$ as well.}. The trapping of the particle happens on a horizontal trench instead of a vertical one.
This results in a perfect Andreev reflection at $E \rightarrow 0$, i.e., $P(G) = \delta(G+G_Q)$. The above distribution functions
identify $\lambda > 0$ and $\lambda < 0$ as a trivial and topological phases, respectively.

Because of the trapping, the particle completes a cycle in the $(w_1, w_2)$-plane only by rare events of the overbarrier ``thermal'' activation, allowed by the noise term in Eq.~\eqref{eq:langevin}.
The height of the barrier is $\Delta U = 2 |\lambda| \ln (v / E l_0) / l_0$ while the effective temperature equals $1/l_0$ \footnote{The effective temperature coincides with the diffusion coefficient $1/l_0$, consistently with Einstein relation for a Brownian particle with a unit mobility.}.
Then, using Eq.~\eqref{eq:N_E}, we can estimate $N(E) \propto \exp (- l_0 \Delta U) \propto E^{2|\lambda|}$ which yields
\begin{equation}
    \nu(E) \propto \frac{1}{E^{1 - 2|\lambda|}}
\end{equation}
for the DOS. Conclusions for $G(E)$ and $\nu(E)$ hold as long as the activation barrier is large, $\Delta U \gtrsim 1/l_0$, which translates to $E \lesssim  (v/l_0)\exp(-c/|\lambda|)$ with $c\sim 1$. At higher energies $G(E)$ and $\nu(E)$ behave in the same way as at the critical point, $\lambda = 0$.

In fact, condition $\lambda = 0$ arises naturally in the limit of strong spin-orbit coupling, see Eq.~\eqref{eq:crit_condition} and the preceding discussion. 
Under this condition, the particle undergoes a free Brownian motion along the equipotential trenches. 
It completes a cycle in a ``time'' 
$\Delta x \sim l(E)$ with $l(E) = l_0 \ln^2(v / E l_0)$. The length scale $l(E)$ plays the role of a correlation radius at a critical point at energy $E$. As a result, we find $N(E) \sim 1/\Delta x \sim [l_0 \ln^2(v / E l_0)]^{-1}$. Thus, the DOS $\nu(E) = \partial N(E)/\partial E$ has a Dyson singularity \cite{dyson1953} at the Fermi level:
\begin{equation}\label{eq:dos_crit}
    \nu(E) \propto \frac{1}{E \ln^3(v / E l_0)}.
\end{equation}
This singularity is indicative of the topological phase transition in one-dimensional superconductors \cite{motrunich2001, brouwer2011}.

\begin{figure}[t]
  \begin{center}
    \includegraphics[scale = 1]{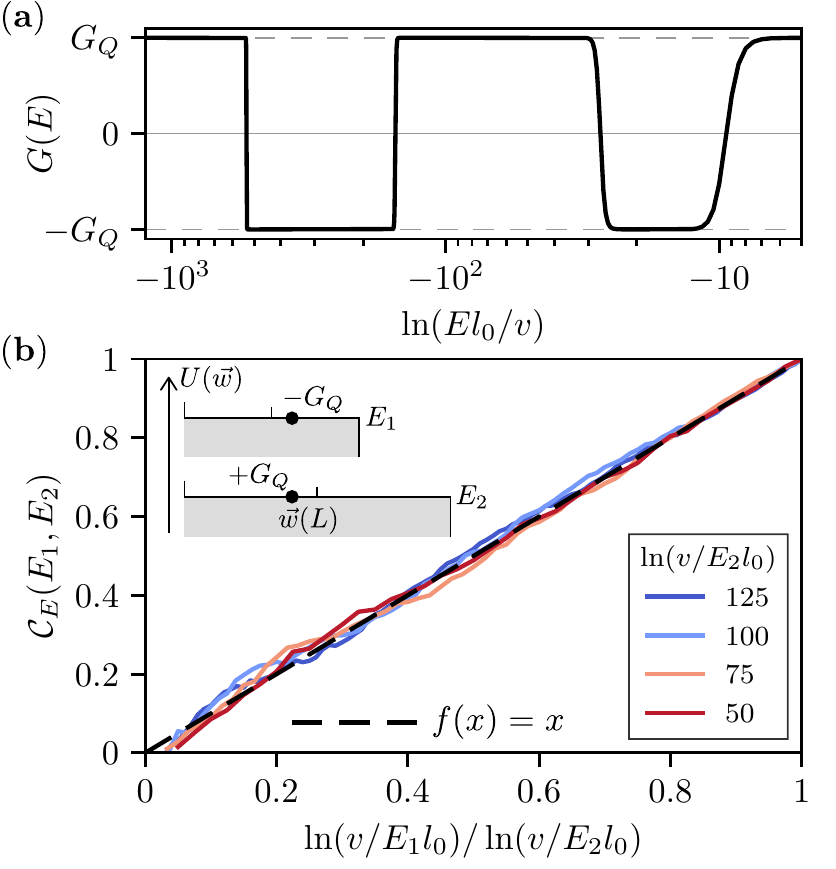}
    \caption{
    Critical behavior of conductance $G(E)$. 
    (a) At low energies, $G(E)$ switches stochastically between the quantized values $\pm G_Q$. Note that the scale is double-logarithmic in $E$. To plot the curve, we numerically simulated the $S$-matrix evolution for $L = 9\cdot 10^4 l_0$ and used Eq.~\eqref{eq:cond}.
    (b)~Correlation function ${\cal C}_E(E_1, E_2)$ of the conductances at different energies (solid lines).
    ${\cal C}_E(E_1, E_2)$ is evaluated numerically; to perform the averaging over disorder realizations in the definition of ${\cal C}_E$, we computed $G(E)$ for $2500$ samples [we use a version of Eq.~\eqref{eq:langevin_big} simplified in the low-energy limit to find $G(E)$, see Eq.~(S43) of \cite{sm}].
    The curves plotted for different $\ln (v / E_2 l_0)$ coincide with each other. This verifies scaling relation~\eqref{eq:scaling}. Scaling function $f(x)$ is well approximated by $f(x) = x$ (dashed line). Inset: Deviation of ${\cal C}_E(E_1, E_2)$ from unity stems from the configurations in which the $\vec{w}$-particles at energies $E_1$ and $E_2$ end their evolution on the opposite halves of the equipotential trenches. 
    }
    \label{fig:scaling}
  \end{center}
\end{figure}

The value of the conductance at the critical point is random and sample specific.
It is determined by the end-point of the particle's random walk along the equipotential trenches, as shown in Fig.~\ref{fig:potential}.
In the limit $E \rightarrow 0$, the conductance distribution function approaches $P(G) = [\delta(G - G_Q) + \delta(G + G_Q)] / 2$.
To understand this result, we note that the critical state can be pictured
as an alternating sequence of trivial- and topological-phase domains \cite{motrunich2001} of a typical size $l(E)$.
Conductance measured
at bias $eV = E$
depends on the type of the domain adjacent to the superconductor's end.
If this domain is topological, then the scattering of an incident electron happens similarly to that off a conventional Majorana wire, $G = -G_Q$ (we remind that $G$ characterizes the backscattered current).
In the opposite case, it is similar to the scattering off an insulator, $G = +G_Q$. The type of the end domain depends on disorder realization, but the two possibilities are equally probable reflecting the criticality.

\sctn{Differential conductance at the critical point.---}To quantify the behavior of $G(E)$ at $\lambda = 0$, we find the conductance correlation function ${\cal C}_E(E_1, E_2) = \langle G(E_1) G(E_2)\rangle / \langle G^2 \rangle$
at $T = 0$.
The conductance at energy $E$ is determined by a disorder realization in a segment of the superconductor of size $\sim l(E)$ adjacent to its end; the correlation radius $l(E)$ is the only relevant length scale at the critical point.
In the spirit of the infinite-randomness model \cite{fisher1994}, we expect that the energy-dependence of the conductance comes only from the respective dependence of $l(E)$. Thus the dimensionless function ${\cal C}_E(E_1, E_2)$ must have a one-parameter scaling form:
\begin{equation}\label{eq:scaling}
    {\cal C}_E (E_1, E_2) = \tilde{f}(l(E_1) / l(E_2)) = f\left(\frac{\ln \frac{1}{E_1}}{\ln \frac{1}{E_2}}\right)
\end{equation}
(hereinafter we suppress $v / l_0$ under the logarithms for brevity).
Without loss of generality, we will assume below that $0 < E_2 < E_1$, such that the argument of scaling function $f(x)$ satisfies $0 < x < 1$.

The form of $f(x)$ can be established analytically for $1 - x \ll 1$.
To do that, we compare the results of the evolution of the $\vec{w}$-particles at two close energies $E_1$ and $E_2$.
While the noise acting on $\vec{w}_{E_1}$ and $\vec{w}_{E_2}$ is the same [see Eq.~\eqref{eq:langevin_big}], the potential landscapes in which they move are different: the lengths of the respective trenches differ by a relative amount $1 - \ln \frac{1}{E_1} / \ln \frac{1}{E_2}$, see Eq.~\eqref{eq:potential}.
The conductances $G(E_1)$ and $G(E_2)$ are opposite to each other if the two particles end up on the different halves of the respective trenches, see Fig.~\ref{fig:scaling}. Such configurations reduce the correlation function from unity.
Let us consider a single cycle of motion of the $\vec{w}$-particles.
$\vec{w}_{E_1}$ and $\vec{w}_{E_2}$ start their motion at the beginning of the respective trenches and move synchronously.
However, they reach the middles of their trenches at different ``times'', as the trenches have different lengths.
It takes a typical time $\Delta x \sim l(E)$ for a particle to reach the middle of the trench. At that time, the probability density to find the particle at distance $\delta w$ from the beginning of the trench is given by the diffusion kernel $P(\delta w, \Delta x) = (\pi \Delta x / l_0)^{-1/2} \exp\bigl(- \frac{\delta w^2}{4 \Delta x / l_0}\bigr)$. We can estimate $1 - {\cal C}_E(E_1, E_2)$ as the probability for two particles to be on different sides of the respective middle points:
\begin{align}
    1 - {\cal C}_E(E_1, E_2) \sim  &\int_0^{2\ln\frac{1}{E}} d(\delta w)\Bigl[1 - {\rm sign}\bigl(\delta w - \ln \tfrac{1}{E_1}\bigr)\notag \\
    &\times {\rm sign}\bigl(\delta w - \ln \tfrac{1}{E_2}\bigr) \Bigr] P(\delta w, \Delta x).
\end{align}
By estimating the integral, we find:
\begin{equation}\label{eq:corr_close_energies}
    1 - {\cal C}_E(E_1, E_2) \sim 1 - \frac{\ln \tfrac{1}{E_1}}{\ln \tfrac{1}{E_2}},
\end{equation}
i.e., $1 - f(x) \sim 1 - x$ for $1 - x \ll 1$.

For an arbitrary relation between $E_1$ and $E_2$, we find the correlation function numerically by directly simulating Eq.~\eqref{eq:langevin_big} (see \cite{sm} for details). The result of the simulation 
is presented in Fig.~\ref{fig:scaling}(b). Surprisingly, $f(x)$ appears to be well approximated by $f(x) = x$ in the whole interval $x \in [0, 1]$.

Finally, Fig.~\ref{fig:scaling}(a) demonstrates $G(E)$ for a particular realization of the disorder.
At low energies, the values of the conductance stochastically alternate between $+G_Q$ and $-G_Q$. These changes are
uniform in variable $\ln\ln (1/E)$, cf.~Eq.~\eqref{eq:scaling}. It means that the fluctuations of $G(E)$ become increasingly dense at $E\to 0$. At the lower end, the rapid fluctuations are cut off either by energy $E_L \sim (v / l_0) \exp( - \tilde{c} \sqrt{L / l_0})$ at which the correlation radius $l(E)$ becomes comparable to the system size $L$ (with $\tilde{c} \sim 1$), or by the level broadening $\Gamma$ induced by vortices.

The $\ln\ln E$ scale comes from the mechanism of the conductance variations. Upon the decrease of energy, $G(E)$ jumps from $G_Q$ to $-G_Q$ every time a new energy level appears on the length scale $l(E) \sim l_0 \ln^2 (1/E)$. The respective change of energy satisfies the condition $\nu(E) l(E) \delta E \sim 1$, which can be cast in the form $\delta [\ln \ln (1/E)] \sim 1$ with the help of Eq.~\eqref{eq:dos_crit} for the DOS.
The emergence of the double-logarithmic scale was also noticed in Ref.~\onlinecite{shivamoggi2010}.

\begin{figure}[t]
  \begin{center}
    \includegraphics[scale = 1]{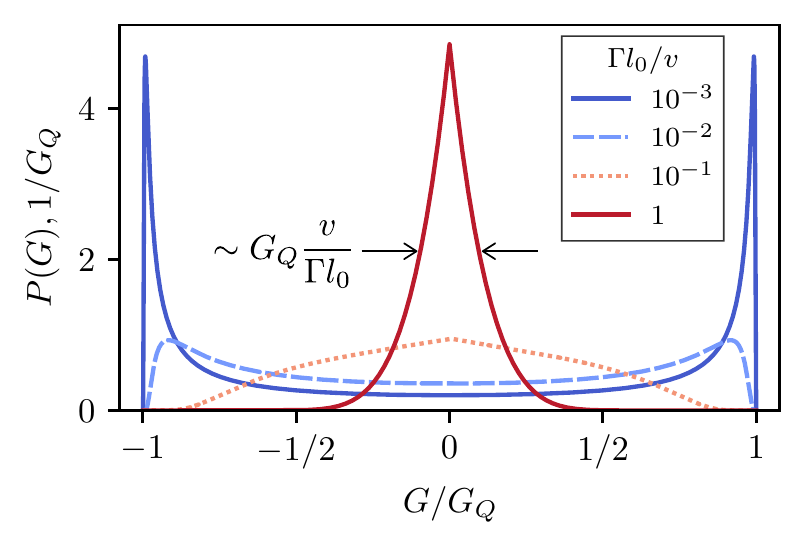}
    \caption{Distribution function $P(G)$ of the zero-bias conductance in the presence of the quasiparticle absorption induced by the vortices. When the absorption is weak, $\Gamma l_0 / v \ll 1$, the conductance is close to one of the two quantized values $\pm G_Q$. Stronger absorption reduces the characteristic magnitude of $G$. For $\Gamma l_0 / v \gtrsim 1$, the conductance is symmetrically distributed in a narrow interval around $G = 0$, see Eq.~\eqref{eq:distribution}. An expression for $P(G)$ valid at arbitrary $\Gamma$ is presented in the supplement \cite{sm}.}
    \label{fig:distribution}
  \end{center}
\end{figure}

\sctn{Influence of vortices.---}
So far, we have neglected the influence of vortices on the conductance. 
An incident electron may sink into the vortex core thus dropping out of the backscattered current. Thus vortices suppress the magnitude of $G$.
To illustrate this effect, we find the distribution function $P(G)$ in the regime of strong absorption, $\Gamma \gg v / l_0$.
In this regime, an incident electron undergoes at most a single EC or CAR process.
This allows us to find the respective amplitudes perturbatively in $\eta_{\rm A / N}(x)$. Using Eq.~\eqref{eq:langevin_S}, we obtain at $E \ll \Gamma$ \footnote{Forward scattering phases $\vartheta_{R/L}$ are inconsequential for finding $P(G)$ so we suppress them in Eq.~\eqref{eq:A_AN}.}:
\begin{equation}\label{eq:A_AN}
    A_{\rm A / N} = -\frac{i}{2}\int_{0}^{L}dx\,e^{- \frac{2\Gamma}{v} (L - x)} \eta_{\rm A / N}(x).
\end{equation}
The conductance distribution function can be expressed as $P(G) = \langle \delta(G - G_Q(|A_{\rm N}|^2 - |A_{\rm A}|^2))\rangle$, where $\langle \dots \rangle$ denotes the average over the realizations of $\eta_{\rm A / N}(x)$. Substituting Eq.~\eqref{eq:A_AN} into this expression and using Eq.~\eqref{eq:correlators}, we obtain at the critical point:
\begin{equation}\label{eq:distribution}
P(G) = \frac{1}{G_Q}\frac{4\Gamma l_0}{v} \exp \Bigl[- \frac{8\Gamma l_0}{ v} \frac{|G|}{G_Q}\Bigr].    
\end{equation}
The absorption renders the typical magnitude of the conductance small, $|G| \sim G_Q (v / \Gamma l_0) \ll G_Q$. This is in contrast to the case of $\Gamma = 0$, in which $G = \pm G_Q$. 
Importantly, $P(G)$ remains symmetric at all values of $\Gamma$ reflecting the criticality.
The crossover between a single-peak $P(G)$ at $\Gamma \gg v / l_0$ and its two-peak counterpart in the opposite limit is illustrated in Fig.~\ref{fig:distribution}.
The regime of strong quasiparticle absorption may be relevant for the recent experiments \cite{lee2017, gul2021}, in which $|G| \ll G_Q$ was measured.

\sctn{Parametric correlations.}---In the conventional mesoscopic transport, there is no need to collect the data from an ensemble of devices to measure the statistics of the conductance fluctuations. The ensemble averaging can be achieved in a \textit{single} sample by varying its parameters, such as the electron density \cite{lee1985, altshuler1985, lee1987}. This result is known as the ergodicity hypothesis.

The variation $\delta n$ of the electron gas density changes the Fermi momentum of edge states electrons by $\delta k_\mu = \delta n  (\partial \mu / \partial n) / v$, where $\partial \mu / \partial n$ is the inverse compressibility of the quantum Hall state.
The change of $k_\mu$ affects the phases of the EC amplitudes, $\eta_{\rm N}(x) \rightarrow \eta_{\rm N}(x) e^{-2i\delta k_\mu x}$, see Eq.~\eqref{eq:EC_basic}.
However, the CAR amplitudes remain approximately intact, $\eta_{\rm A}(x) \rightarrow \eta_{\rm A}(x)$, see Eq.~\eqref{eq:CAR_basic}.
Because of this, the ergodicity may appear to \textit{break down} for the proximitized counter-propagating edges.
The variation of $\eta_{\rm A}(x)$ with $k_\mu$ happens due to subtle effects only, such as the non-local character of the electron tunneling between the edges. Indeed, the $x$-coordinate of a tunneling electron may change by $\sim \sqrt{\xi d}$ which sets a scale $\sim 1 / \sqrt{\xi d}$ for the variation of $\eta_{\rm A}(x)$ with $k_\mu$.

To demonstrate the existence of two scales for the density variation, we compute the correlation function ${\cal C}_n(\delta n) = \langle G(n + \delta n) G(n) \rangle / \langle G^2 \rangle$, focusing on the regime of strong quasiparticle absorption, $\Gamma \gg v / l_0$. In this regime, one can use the perturbative expressions of Eq.~\eqref{eq:A_AN} to find ${\cal C}_n(\delta n)$. Substituting them in Eq.~\eqref{eq:conductance} and using Eq.~\eqref{eq:correlators} (together with its counterpart for the amplitude correlations at different $k_\mu$ \cite{sm}), we obtain:
\begin{equation}\label{eq:corr_function}
    {\cal C}_n(\delta n) = \frac{1}{2}\exp\Bigl[-\Bigl(\frac{\delta n}{n_{\rm c, A}}\Bigr)^2\Bigr] + \frac{1}{2} \frac{1}{1 + (\delta n / n_{\rm c, N})^2}.
\end{equation}
Here 
$n_{\rm c, N} = 2\Gamma (\partial n / \partial \mu)$ and $n_{\rm c, A} = (v / \sqrt{\xi d})(\partial n / \partial \mu)$ are the two scales of the variation of $G$ with $n$.
We see that the latter of the two scales diverges at $\sqrt{\xi d} \rightarrow 0$ leading to the saturation of ${\cal C}_n(\delta n)$ at $\delta n \gg  n_{\rm c, N}$ and creating an appearance that the ergodicity breaks down.

\begin{figure}[t]
  \begin{center}
    \includegraphics[scale = 1]{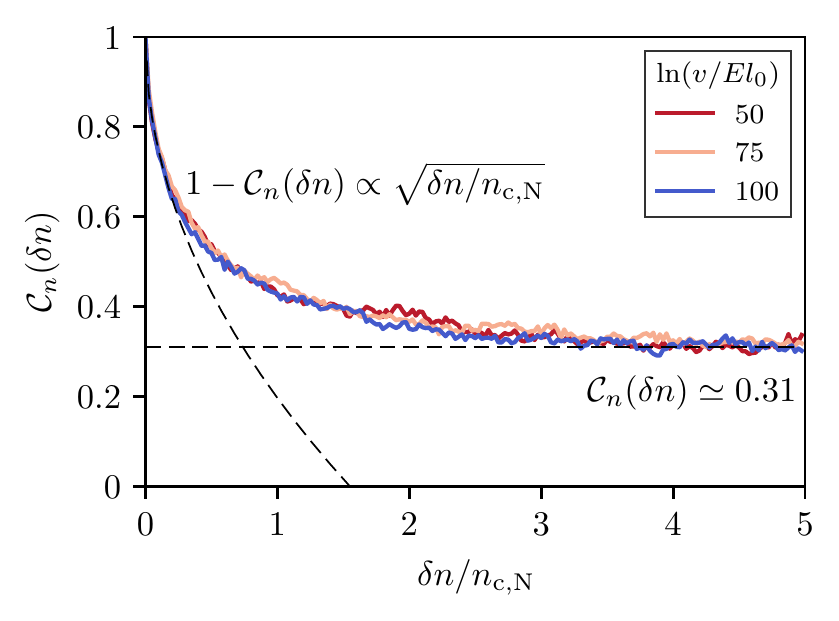}
    \caption{Correlation function ${\cal C}_n(\delta n)$ of the conductances at different electron densities for $\delta n \ll n_{\rm c, A}$. We compute ${\cal C}_n(\delta n)$ by simulating the evolution of the $S$-matrix for $L = 9\cdot 10^4 l_0$ [see Eq.~\eqref{eq:langevin_big} and its low-energy counterpart, Eq.~(S34) of \cite{sm}], and averaging the result over $5000$ samples. The saturation of ${\cal C}_n(\delta n)$ stems from the insensitivity of the CAR amplitudes to the density variations $\delta n \ll n_{\rm c, A}$. The curves plotted for different $\ln(v / E l_0)$ coincide with each other.}
    \label{fig:density_corr}
  \end{center}
\end{figure}

The saturation of ${\cal C}_n(\delta n)$ persists at smaller values of~$\Gamma$. In the limit $\Gamma \rightarrow 0$, the parameter $\Gamma$ in the scale $n_{\rm c, N}$ is replaced by $v / l_0$.
To see the saturation at $\Gamma = 0$, we find ${\cal C}_n(\delta n)$ by numerically simulating Eq.~\eqref{eq:langevin_big}.
The result of the simulation is presented in Fig.~\ref{fig:density_corr} for different values of $E$.
We see that ${\cal C}_n(\delta n)$ saturates at $\simeq 0.31$ independent of $E$.
This can be explained by analyzing the Fokker-Planck equation for the joint distribution function of the $\vec{w}$-variables at two values of density separated by $\delta n \gg n_{\rm c, N}$ \cite{sm}. In this limit, the Fokker-Planck equation 
acquires an elliptic form. The independence of ${\cal C}_n(\delta n)$ of $E$ stems from its scaling properties.

In fact, ${\cal C}_n(\delta n)$ plotted for different energies collapse on the same curve not only at $\delta n \gg n_{\rm c, N}$, but at all values of $\delta n$.
To substantiate this observation, we find ${\cal C}_n(\delta n)$ analytically at $\delta n \ll n_{\rm c, N}$.
The shift $\delta k_\mu \propto \delta n$ 
leads to the imperfect correlation between the noises acting on the $\vec{w}$-particles at the two values of density.
As a result, the particles separate in the course of their motion, $|\vec{w}_n - \vec{w}_{n + \delta n}| \sim \sqrt{\delta k_\mu \Delta x}$, where $\Delta x$ is the ``time'' measured from the start of the last cycle \cite{sm}.
Due to the separation, they may end up on the opposite halves of the respective trenches leading to the deviation of ${\cal C}_n(\delta n)$ from unity.
The motion of an individual particle happens with the diffusion coefficient $1/l_0$. It takes time 
$\Delta x \sim l(E)$ for the particles to spread over the respective trenches (which have lengths $\sim\sqrt{l(E)/l_0}$). The characteristic particle separation at that time is $\sim\sqrt{\delta k_\mu l(E)}$. Given the characteristic particle density along a trench $1/\sqrt{l(E)/l_0}$, the probability to find the two particles at opposite sides of the respective trench middle points is $\sim\sqrt{\delta k_\mu l(E)}/\sqrt{l(E)/l_0}=\sqrt{\delta k_\mu l_0}$. Thus, we find $1 - {\cal C}_n(\delta n) \sim \sqrt{\delta n / n_{\rm c, N}}$ with $n_{\rm c, N} = (v / l_0) (\partial n / \partial \mu)$. The square-root behavior of ${\cal C}_n(\delta n)$ at $\delta n \ll n_{\rm c, N}$ and its independence of $E$ is in agreement with the numerical calculation.

Our theory offers an interpretation of the observations of Refs.~\onlinecite{lee2017} and \onlinecite{gul2021}.
In these experiments, a dirty superconductor provided coupling between two $\nu = 1$ counter-propagating edges. The basic assumptions of our model are consistent with this setup.
Therefore, we expect the devices to be at the critical point between the topological and trivial phases, and the conductance distribution function to be symmetric, $P(G) = P(-G)$.
In general, $P(G)$ can be found by sampling $G$ in a given device by varying its electron density $n$.
However, Refs.~\onlinecite{lee2017} and \onlinecite{gul2021} reported a negative conductance weakly sensitive to the variations of $n$.
We can explain this observation by a large correlation scale $n_{\rm c, A}$ of CAR processes.

Since the measured conductance $|G| \ll G_Q$, we focus on the regime of strong quasiparticle absorption by vortices, $\Gamma \gg v / l_0$. In this regime, the width of $P(G)$ is $\sim G_Q(v / \Gamma l_0)$.
The conductance can be related to the probabilities $p_{\rm N}$ and $p_{\rm A}$ of EC and CAR processes as $G = G_Q(p_{\rm N} - p_{\rm A})$.
In the considered perturbative regime, $p_{\rm N}$ and $p_{\rm A}$ are independent of each other, and have typical values $\sim v / \Gamma l_0$.
The former probability varies with $n$ on the scale $n_{\rm c, N}$, while the latter one changes on a much larger scale $n_{\rm c, A}$.
Suppose that the accessible measurement range of density $n$ is smaller than $n_{\rm c, A}$ but exceeds $n_{\rm c, N}$.
Then, $p_{\rm A}$ stays approximately constant within the range while $p_{\rm N}$ fluctuates with variation of $n$.
The statistics of conductance collected in such a measurement would be $\tilde{P}(G) = \Theta(G + G_Q p_{\rm A}) \frac{8\Gamma l_0}{v}\exp\bigl[- \tfrac{8\Gamma l_0}{v} \tfrac{G + G_Q p_{\rm A}}{G_Q}\bigr]$.
Accordingly, the probability to measure a negative conductance is $1 - \exp[ -(8\Gamma l_0 / v) p_{\rm A}]$. It is close to $1$ if $p_{\rm A}$ is relatively large, $p_{\rm A} \gtrsim v / (8\Gamma l_0)$.
With an assumption of an anomalously large $p_{\rm A}$, this may explain the negative signal reported in Ref.~\onlinecite{gul2021}, and even without such an assumption the data of Ref.~\onlinecite{lee2017}.
A similar mechanism may be relevant for the observations at other integer fillings~$\nu$.

\sctn{Conclusions.}---Transport of a quantum Hall edge across a narrow superconductor is determined by the competition of CAR and EC processes.
For a disordered superconductor, amplitudes of these processes are random but are balanced statistically,
see Eqs.~\eqref{eq:CAR_EC_short} and \eqref{eq:CAR_EC_final}.
The balance automatically tunes the system to the critical point between trivial and topological phases.
The charge transport at the critical point is random.
At low bias $V = E/ e$ (see Fig.~\ref{fig:setup} for the setup), conductance $G(E)$ is equally distributed between two quantized values, $\pm G_Q$. 
Which value of $G$ is realized at a given $E$ is determined by the disorder configuration in a segment of superconductor of length $l(E) \propto \ln^2 (1 / E)$.
Upon changing $E$, $G(E)$ switches stochastically between $\pm G_Q$, see Fig.~\ref{fig:scaling}(a).
The switchings are roughly equidistant in $\ln \ln (1/E)$ scale, see Eq.~\eqref{eq:scaling} and Fig.~\ref{fig:scaling}.
Electron tunneling into the vortex cores breaks the quantization of $G$. 
A strong quasiparticle loss shrinks the conductance distribution $P(G)$ to a narrow interval of values around $G = 0$, see Eq.~\eqref{eq:distribution} and Fig.~\ref{fig:distribution}.
$P(G)$ can be determined  experimentally by collecting the statistics of the conductance fluctuations with electron density $n$. 
To achieve the representative sampling of $G$, the variation of $n$ has to exceed the scale $n_{\rm c, A}$ at which the CAR amplitudes change. 
At smaller density variations, the conductance may appear non-ergodic, see Eq.~\eqref{eq:corr_function} and Fig.~\ref{fig:density_corr}.
Such a seemingly non-ergodic behavior may be directly relevant for the data of Refs.~\onlinecite{lee2017, gul2021}.

Our theory identifies a challenge in engineering a topological superconductor by proximity-coupling the quantum Hall edges.
It demonstrates that one \textit{cannot} reach a topological phase when using a dirty superconductor
to induce the proximity effect, even when the spin-orbit interaction is strong.
At the same time, it shows that the proximity-coupled edges give an unprecedented access to the fundamentally interesting physics of the topological phase transition criticality. There is no need for fine-tuning of the magnetic field or the chemical potential because the device self-tunes to the critical point naturally.
Looking forward, it would be interesting to extend our theory to the case of the proximity-coupled fractional quantum Hall edges.

\textit{Acknowledgements}.---
We acknowledge very useful discussions with Pavel D.~Kurilovich and Gil Refael. V.D.K. also acknowledges an advice of Artem Merezhnikov on optimization of the numeric calculations.
This work is supported by
NSF DMR-2002275 (V.D.K.) and ARO W911NF2210053 (L.I.G.).

\bibliography{references}

\end{document}


\title{Supplemental Information for ``Criticality in the crossed Andreev reflection of a quantum Hall edge''}

\author{Vladislav D.~Kurilovich}
\affiliation{Department of Physics, Yale University, New Haven, CT 06520, USA}
\author{Leonid I.~Glazman}
\affiliation{Department of Physics, Yale University, New Haven, CT 06520, USA}

\maketitle

\onecolumngrid
\vspace*{-10mm}

\section{Derivation of $1/\lA$ and $1/\lN$, Eq.~(6)\label{sec:lAlN}}
In this section, we present details of the derivation of Eq.~(6) for $1/l_{\rm A}$ and $1/l_{\rm N}$. We start with the expressions for the amplitudes of the crossed Andreev reflection (CAR) and elastic cotunneling (EC), which were derived in the main text, see Eq.~(4). We also present them here for convenience:
\begin{equation}\label{eq:amplitudes}
    \delta A_{\rm A}\hspace{-0.1cm} =\hspace{-0.07cm} \frac{t^2}{v}\hspace{-0.1cm}\int_x^{x+\delta L}\hspace{-0.3cm}  dxdx^\prime e^{-ik_\mu(x^\prime - x)} \partial^2_{yy^\prime}{\cal G}_{\rm he}^{\downarrow\uparrow}(x, 0;x^\prime, d | E = 0),\quad
    \delta A_{\rm N}\hspace{-0.1cm} = \hspace{-0.07cm} \frac{t^2}{v}\hspace{-0.1cm} \int_x^{x+\delta L}\hspace{-0.3cm}  dxdx^\prime e^{-ik_\mu(x^\prime + x)} \partial^2_{yy^\prime}{\cal G}_{\rm ee}^{\uparrow\uparrow}(x, 0;x^\prime, d | E = 0),
\end{equation}
where $x$ is the $x$-coordinate of the considered short element.
Finding $\delta A_{\rm A/ N}$ requires the knowledge of the superconductor Green's function (we made the energy argument of ${\cal G}$ explicit; in what follows, we assume that the two-dimensional electron gas is situated at $z = 0$ and take $z, z^\prime = 0$ in all of the Green's functions).
For simplicity, we begin by considering a type I superconductor. Such a superconductor does not admit the magnetic field, so it is possible to choose the gauge in which the vector potential vanishes in the superconductor and the order parameter is real. 
We describe the superconductor by the following BCS Hamiltonian:
\begin{equation}\label{eq:H_SC}
    H_{\rm SC} = \sum_{\sigma\sigma^\prime}\int d^3r\,\chi^\dagger_\sigma({\bm r})\Big[\Bigl(-\frac{\partial_{\bm r}^2}{2m} - \mu\Bigr)\delta_{\sigma \sigma^\prime} + U_{\sigma \sigma^\prime}({\bm r})\Big] \chi_{\sigma^\prime}({\bm r}) + \int d^3 r\,\Delta \big(\chi_\uparrow^\dagger ({\bm r})\chi_\downarrow^\dagger({\bm r}) + \chi_\downarrow ({\bm r})\chi_\uparrow ({\bm r})\big).
\end{equation}
Here $\chi_{\sigma} (\bm{r})$ is an annihilation operator for an electron with spin $\sigma$, $\mu$ is the chemical potential, $m$ is the effective mass, and $\Delta$ is the superconducting order parameter. Term $U_{\sigma \sigma^\prime}(\bm{r})$ includes both the disorder potential $U_0(\bm{r})$ and the spin-orbit interaction associated with it. In the momentum representation,
\begin{equation}
    U_{\sigma \sigma^\prime}(\bm{k}, \bm{k}^\prime) = U_0(\bm{k} - \bm{k}^\prime) \bigl(\delta_{\sigma \sigma^\prime} + i \gamma\,\bm{\sigma}_{\sigma \sigma^\prime}\cdot[\bm{k} \times \bm{k}^\prime]/p_F^2\bigl),
\end{equation}
where $\gamma$ is a dimensionless parameter characterizing the strength of the spin-orbit interaction; $p_F$ is the Fermi momentum in the metal. We assume that the spin-orbit scattering is weak compared to the spin-conserving one, $\gamma \ll 1$, and model the disorder potential as a Gaussian random variable with a short-ranged (i.e., momenta independent) correlator, $\langle |U_0(\bm{k} - \bm{k}^\prime)|^2 \rangle =(2\pi \nu_{\rm M} \tau_{\rm mfp})^{-1}$.
We expressed the correlator in terms of the normal-state density of states in the metal $\nu_{\rm M}$ and the electron mean free time $\tau_{\rm mfp}$. In what follows, we focus on the limit of a ``dirty'' superconductor, $\Delta \tau_{\rm mfp} \ll 1$.

To find $1/l_{\rm A}$ and $1 / l_{\rm N}$, we compute the variances of the CAR and EC amplitudes starting with the former one. Using Eq.~\eqref{eq:amplitudes}, we can represent $\langle |\delta A_{\rm A}|^2 \rangle$ as
\begin{equation}\label{eq:var1}
        \langle |\delta A_{\rm A}|^2 \rangle = \frac{t^4}{v^2}\int^{x+\delta L}_{x}\hspace{-0.3cm}
        dx_1 dx_1^\prime dx_2 dx_2^\prime\,
        e^{ik_\mu(x_1  - x_1^\prime - x_2 +x_2^\prime)} \partial^2_{y_1,y_1^\prime}\partial^2_{y_2, y_2^\prime}\Lngl {\cal G}^{\downarrow \uparrow}_{\rm he}(\bm{r}_1 ; \bm{r}_1^\prime | E = 0)\cdot {\cal G}^{\uparrow \downarrow}_{\rm eh}(\bm{r}_2^\prime ; \bm{r}_2 | E = 0)\Rngl\Bigr|^{y_1^\prime, y_2^\prime = d}_{y_1, y_2 = 0}.
\end{equation}
Note that on the right hand side we replaced the average with its irreducible component. We neglect a contribution $\propto \langle {\cal G}\rangle \langle {\cal G}\rangle$ in Eq.~\eqref{eq:var1} due to the exponential smallness of the disorder-averaged Green's function,  $\langle {\cal G}\rangle \propto \exp (-d / 2l_{\rm mfp})$ ($l_{\rm mfp}$ is the electron mean free path).
It is convenient to relate the Green's functions of the superconductor to the retarded and advanced Green's function of the metal in the normal state. The relation reads:
\begin{equation}
    {\cal G}_{\tau \tau^\prime}^{\sigma \sigma^\prime}({\bm r}; {\bm r}^\prime| E) = \int \frac{d\epsilon}{2\pi i} \frac{[E + \epsilon \tau_z + \Delta \tau_x]_{\tau \tau^\prime}}{\Delta^2 - E^2 + \epsilon^2}\Bigl[\GR_{\rm N}({\bm r}; {\bm r}^\prime|\epsilon) - \GA_{\rm N} ({\bm r}; {\bm r}^\prime|\epsilon)\Bigr]_{\sigma \sigma^\prime},
\end{equation}
where $\tau_{x,z}$ are the Pauli matrices in the Nambu space. With its help, we obtain
\begin{align}
    &\langle |\delta A_{\rm A}|^2\rangle = \frac{t^4}{4\pi^2 v^2} \int_x^{x+\delta L}\hspace{-0.3cm}dx_1 dx_1^\prime dx_2 dx_2^\prime e^{ik_\mu(x_1 - x_1^\prime - x_2 +x_2^\prime)} \int \frac{\Delta d\epsilon_1}{\Delta^2 + \epsilon^2_1} \frac{\Delta d\epsilon_2}{\Delta^2 + \epsilon^2_2} \notag\\
    &\times \partial^2_{y_1,y_1^\prime}\partial^2_{y_2,y_2^\prime}
    \Bigl[\Lngl \GR_{\rm N,\downarrow\uparrow}(\bm{r}_1;\bm{r}_1^\prime|\epsilon_1)\cdot \GA_{\rm N, \uparrow \downarrow}(\bm{r}_2^\prime;\bm{r}_2|\epsilon_2)\Rngl + \Lngl \GR_{\rm N, \uparrow\downarrow}(\bm{r}_1^\prime;\bm{r}_1|\epsilon_1)\cdot \GA_{\rm N, \downarrow\uparrow}(\bm{r}_2;\bm{r}_2^\prime|\epsilon_2)\Rngl \Bigr]_{y_1,y_2 = 0}^{y_1^\prime, y_2^\prime = d}.\label{eq:var2}
\end{align}
In arriving to Eq.~\eqref{eq:var2}, we dispensed with terms $\lngl {\cal G}_{\rm N}^{\rm R} \cdot {\cal G}_{\rm N}^{\rm R} \rngl$ and $\lngl {\cal G}_{\rm N}^{\rm A} \cdot {\cal G}_{\rm N}^{\rm A} \rngl$. 
Combinations ${\cal G}_{\rm N}^{\rm R} {\cal G}_{\rm N}^{\rm R}$ and ${\cal G}_{\rm N}^{\rm A}{\cal G}_{\rm N}^{\rm A}$ rapidly oscillate with their positional arguments (on a scale set by the Fermi wavelength $\lambda_{F}$ in the metal), and therefore produce a negligible contribution to $\langle |\delta A_{\rm A}|^2\rangle$.
The averages of the form $\lngl {\cal G}^{\rm R}_{\rm N} \cdot {\cal G}^{\rm A}_{\rm N} \rngl$ can be expressed in terms of the normal-state diffuson. Following a standard procedure \cite{hek94, abrikosov}, we find to the leading order in a small parameter $\lambda_F / l_{\rm mfp}\ll 1$:
\begin{align}
    &\Lngl \GR_{\rm N,\sigma_1 \sigma_1^\prime}({\bm r}_1;{\bm r}_1^\prime|\epsilon_1)\cdot \GA_{\rm N, \sigma_2^\prime \sigma_2}({\bm r}_2^\prime; {\bm r}_2|\epsilon_2)\Rngl\notag\\
    &= \frac{1}{2\pi\nu_{\rm M}\tau_{\rm mfp}^{2}}\int d^3rd^3r^\prime\,\langle {\cal G}^{\rm R}_{\rm N}({\bm r}_{1};{\bm r}|\epsilon_1)\rangle\langle {\cal G}^{\rm A}_{\rm N}({\bm r};{\bm r}_{2}|\epsilon_2)\rangle {\cal D}_{\sigma_1 \sigma_2, \sigma_1^\prime \sigma_2^\prime} ({\bm r};{\bm r}^\prime|\epsilon_1-\epsilon_2)\langle {\cal G}^{\rm A}_{\rm N}({\bm r}_2^\prime;{\bm r}^\prime|\epsilon_2)\rangle\langle {\cal G}^{\rm R}_{\rm N}({\bm r}^\prime;{\bm r}_{2}|\epsilon_1)\rangle\label{eq:diffuson}
\end{align}
Here ${\cal D}_{\sigma_1\sigma_2,\sigma_1^\prime\sigma_2^\prime}(\bm{r};\bm{r}^\prime|\epsilon)$ is the diffuson. In addition to its dependence on $\bm{r}$ and $\bm{r}^\prime$, the diffuson is a $4 \times 4$ matrix in the spin space, whose structure is determined by the spin-orbit scattering. We will discuss shortly how this matrix can be found. In Eq.~\eqref{eq:diffuson} we disregarded the contribution of a Cooperon. This contribution is exponentially small in $d / l_{\rm mfp} \gg 1$ in the relevant case of $\bm{r}_{1,2}$ and $\bm{r}_{1,2}^\prime$ being on the opposite sides of the superconducting electrode. 

The average Green's functions in Eq.~\eqref{eq:diffuson} decay on the length scale $\sim l_{\rm mfp}$. On the other hand, the diffuson varies on a much larger scale $\sim \xi$ ($\xi$ is the superconducting coherence length; $\xi \gg l_{\rm mfp}$ for a ``dirty'' superconductor). The separation of scales allows us to approximate $\bm{r}\approx \bm{r}_1$ and $\bm{r}^\prime \approx \bm{r}_2$ in the argument of ${\cal D}_{\sigma_1 \sigma_2, \sigma_1^\prime \sigma_2^\prime}$ in Eq.~\eqref{eq:diffuson}. The remaining integrals over $\bm{r}$ and $\bm{r}^\prime$ can be carried out straightforwardly. Similarly to Ref.~\onlinecite{us}, we find for a combination entering Eq.~\eqref{eq:var2}:
\begin{align}
    \partial^2_{y_1,y_1^\prime}\partial^2_{y_2,y_2^\prime}\Lngl \GR_{\rm N,\downarrow\uparrow}(\bm{r}_1;\bm{r}_1^\prime|\epsilon_1)\cdot \GA_{\rm N, \uparrow \downarrow}(\bm{r}_2^\prime;\bm{r}_2|\epsilon_2)&\Rngl\Bigr|_{y_1,y_2 = 0}^{y_1^\prime, y_2^\prime = d} \notag\\
    = &2\pi \nu_{\rm M} (\pi p_F)^2 \delta(x_1 -x_2)\delta(x_1^\prime - x_2^\prime) {\cal D}_{\downarrow\downarrow, \uparrow\uparrow}(x_1, 0; x_1^\prime, d|\epsilon_1 - \epsilon_2).
\end{align}
Applying the same treatment to the second term in the square brackets in Eq.~\eqref{eq:var2}, and computing the integrals over $\epsilon_{1,2}$ we find for $\langle |\delta A_{\rm A}|^2\rangle$:
\begin{align}\label{eq:var3}
    \langle |\delta A_{\rm A}|^2\rangle = \frac{\pi^3\nu_{\rm M} p_F^2 t^4}{2 v^2} \int_x^{x + \delta L}\hspace{-0.3cm} dx_1 dx_1^\prime\hspace{-0.1cm}\int_0^{+\infty} d\tau e^{-2\Delta \tau}\bigl[{\cal D}_{\downarrow\downarrow,\uparrow\uparrow}(x_1, 0;x_1^\prime, d|\tau) + {\cal D}_{\uparrow\uparrow, \downarrow\downarrow}(x_1^\prime, d; x_1, 0|\tau)\bigr].
\end{align}
In this expression, we converted the diffuson from the energy-domain into the time-domain.

An expression similar to Eq.~\eqref{eq:var3} can also be obtained for the variance of the EC amplitudes. By repeating the steps leading to Eq.~\eqref{eq:var3}, we obtain:
\begin{align}\label{eq:ec}
    \langle |\delta A_{\rm N}|^2\rangle = \frac{\pi^3 \nu_{\rm M} p_F^2 t^4}{2 v^2} \int_x^{x + \delta L}\hspace{-0.3cm} dx_1 dx_1^\prime\int_0^{+\infty}d\tau e^{-2\Delta \tau}\bigl[{\cal D}_{\uparrow\uparrow,\uparrow\uparrow}(x_1, 0;x_1^\prime, d|\tau) + {\cal D}_{\uparrow\uparrow, \uparrow\uparrow}(x_1^\prime, d; x_1, 0|\tau)\bigr].
\end{align}
The only difference of this expression from Eq.~\eqref{eq:var3} is in the spin components of the diffuson.

Let us discuss how the diffuson can be found. In a disordered metal, ${\cal D}_{\sigma_1,\sigma_2,\sigma_1^\prime \sigma_2^\prime}$ satisfies a version of the diffusion equation which is appropriately modified to account for the spin-orbit scattering \cite{vavilov}:  
\begin{align}\label{eq:diff_eq}
    \Bigl([\partial_\tau - D \partial_{\bm r}^2]\delta_{\sigma_1 \zeta_1}\delta_{\sigma_2\zeta_2} +\frac{2}{3\tau_{\rm so}}\zeta_{\sigma_1\sigma_2,\zeta_1\zeta_2}\Bigr) {\cal D}_{\zeta_1\zeta_2,\sigma_1^\prime\sigma_2^\prime}(\bm{r};\bm{r}^\prime|\tau) &= \delta(\tau)\delta(\bm{r} - \bm{r}^\prime)\delta_{\sigma_1 \sigma_1^\prime}\delta_{\sigma_2 \sigma_2^\prime}.
\end{align}
Here $D$ is the diffusion coefficient and $\tau_{\rm so} = 3\tau_{\rm mfp} / (2\gamma^2)$ is the spin-orbit scattering time. Matrix $\zeta$ is given by
\begin{equation}
    \zeta_{\sigma_1\sigma_2,\zeta_1\zeta_2} = \frac{3}{2}\delta_{\sigma_1\zeta_1}\delta_{\sigma_2\zeta_2}  - \frac{1}{2}\bm{\sigma}_{\sigma_1\zeta_1}\cdot\bm{\sigma}_{\zeta_2\sigma_2} = 
    \begin{pmatrix} 
     1 & 0 & 0 & -1\\
     0 & 2 & 0 & 0\\
     0 & 0 & 2 & 0\\
     -1 & 0 & 0 & 1
     \end{pmatrix}.
\end{equation}
The boundary condition for Eq.~\eqref{eq:diff_eq} is $\bm{n}\cdot\partial_{\bm{r}}{\cal D}_{\sigma_1\sigma_2,\sigma_1^\prime\sigma_2^\prime} = 0$, where $\bm{n}$ is the normal to the surface of the metal. This condition corresponds to the vanishing of the probability- and spin-currents through the metal surface. 

Using Eq.~\eqref{eq:diff_eq}, we can represent the components of the diffuson needed for finding $\langle |A_{\rm A/N}|^2\rangle$ as
\begin{equation}\label{eq:diff_kernel}
{\cal D}_{\downarrow\downarrow, \uparrow\uparrow}(\bm{r};\bm{r}^\prime|\tau) = \frac{1}{2}\Bigl[1 {-} \exp\Bigl(- \frac{4\tau}{3\tau_{\rm so}}\Bigr)\Bigr]{\cal D}(\bm{r};\bm{r}^\prime|\tau), \quad\quad 
{\cal D}_{\uparrow\uparrow, \uparrow\uparrow}(\bm{r};\bm{r}^\prime|\tau) = \frac{1}{2}\Bigl[1 {+} \exp\Bigl(- \frac{4\tau}{3\tau_{\rm so}}\Bigr)\Bigr]{\cal D}(\bm{r};\bm{r}^\prime|\tau),
\end{equation}
and ${\cal D}_{\uparrow \uparrow,\downarrow\downarrow}(\bm{r};\bm{r}^\prime|\tau) =  {\cal D}_{\downarrow\downarrow, \uparrow \uparrow}(\bm{r};\bm{r}^\prime|\tau)$.
Here, we introduced the diffusion kernel ${\cal D}$, which satisfies a usual diffusion equation, 
$(\partial_\tau - D \partial_{\bm r}^2){\cal D} (\bm{r};\bm{r}^\prime | \tau) = \delta(\tau) \delta(\bm{r} - \bm{r}^\prime)$. The diffusion kernel depends sensitively on the geometry of the electrode at hand. For simplicity, we will assume that both the width $d$ and the thickness $h$ of the electrode exceed the coherence length, $d, h \gg \xi$. In this case, we readily find
\begin{equation}
    {\cal D}(x_1, 0; x_1^\prime, d|\tau) = \frac{4}{(4\pi D \tau)^{3/2}}\exp\Bigl(-\frac{(x_1 - x_1^\prime)^2 + d^2}{4D\tau}\Bigr)
\end{equation}
(we set the $z$-arguments of ${\cal D}$ to $z,z^\prime = 0$, and make them implicit). Substituting this expression in Eqs.~\eqref{eq:var3} and \eqref{eq:ec} and computing the integral over the time variable $\tau$, we obtain
\begin{equation}\label{eq:vars_final}
    \langle |\delta A_{\rm A/N}|^2\rangle = \frac{\pi^2\nu_{\rm M} p_F^2 t^4}{2 v^2 D }\cdot \delta L \int \frac{d(\delta x)}{\sqrt{d^2 + (\delta x)^2}}\Bigl[e^{-\sqrt{d^2 + (\delta x)^2}/\xi}\mp e^{-\sqrt{d^2 + (\delta x)^2}/\xi^\star} \Bigr],\quad \xi  = \sqrt{\frac{D}{2\Delta}},\quad \xi^\star = \frac{\xi}{\sqrt{1 + \frac{4}{3 \Delta \tau_{\rm so}}}}.
\end{equation}
We also assumed that the length of the element $\delta L \gg \xi$, and carried out an integration over the ``center-of-mass'' coordinate $(x_1 + x_1^\prime)/2$. Finally, we evaluate the remaining integral over $x$, and express the result in terms of the conductivity of the metal in the normal state $\sigma = 2e^2\nu_{\rm M} D$ and conductance per unit length of the interface $g = 2\pi^2 G_Q t^2 \nu_{\rm edge} \nu_{\rm M} p_F$ (where $\nu_{\rm edge} = (2\pi v)^{-1}$ is the density of states at the edge). This leads to Eqs.~(5) and (6) of the main text.

In the derivation above, we assumed that the superconductor is of type I.
A type II superconductor is different in that it admits magnetic field $B$.
The field affects the amplitudes
of the quasiparticle tunneling across the superconductor.
We expect, however, that this effect is weak at low fields, $B \ll H_{\rm orb}$, and can thus be neglected.
To estimate $H_{\rm orb}$, we compute $\langle {\cal G} (\bm{r};\bm{r}^\prime | E = 0){\cal G} (\bm{r}^\prime;\bm{r}| E = 0)\rangle$ to the lowest non-vanishing order in $B$. Dispensing for simplicity with spin-orbit interaction, and following a standard approach, we obtain (we assume that the net supercurrent is zero and choose a gauge in which the order parameter is real, $\nabla \varphi (\bm{r}) = 0$):
\begin{align}
    \langle {\cal G}_{\tau \tau^\prime} (\bm{r};\bm{r}^\prime | 0){\cal G}_{\rho^\prime \rho} (\bm{r}^\prime;\bm{r}| 0)\rangle
    &= \nu \Bigl[{\cal D}(\bm{r}; {\bm r}^\prime | 2 i \Delta)+\delta{\cal D}(\bm{r}; {\bm r}^\prime | 2 i \Delta)\Bigr] (\tau_z^{\sigma \sigma^\prime} \tau_z^{\rho^\prime \rho} + \tau_x^{\sigma \sigma^\prime} \tau_x^{\rho^\prime \rho}),
    \notag\\
    \delta{\cal D}(\bm{r}; {\bm r}^\prime | 2 i \Delta)&= \int d\bm{r}_1 {\cal D}(\bm{r}; {\bm r}_1 | 2 i \Delta) \cdot 2\pi D (e{\bm A}(\bm{r}_1)/c)^2 {\cal D}(\bm{r}_1; {\bm r}^\prime | 2 i \Delta).\label{eq:mag_field}
\end{align}
Here ${\cal D}(\bm{r}; {\bm r}^\prime | \xi)$ is the normal-state diffuson and ${\bm A}(\bm{r})$ is the vector potential associated with the field $B$.
At distances $d \equiv |{\bm r} - {\bm r}^\prime| \gtrsim \xi$, the unperturbed diffuson is ${\cal D}(\bm{r}; {\bm r}^\prime | 2 i \Delta)\sim e^{-d / \xi} / (D d)$. Using this estimate to evaluate $\int d{\bm r}_1\dots$ in $\delta {\cal D}$, we find $\delta {\cal D}(\bm{r};\bm{r}^\prime|2i\Delta) \sim e^{-d / \xi} \xi (e A / c)^2 / D$. Here $A \sim B\cdot d$ is the typical value of the vector potential between points ${\bm r}$ and ${\bm r^\prime}$.
We see that $\delta {\cal D}$ can be neglected in comparison with ${\cal D}$ at sufficiently low fields, $B \ll H_{\rm orb}$, where
\begin{equation}
    H_{\rm orb} = \frac{\Phi_0}{\xi^{1/2} d^{3/2}}
\end{equation}
and $\Phi_0$ is the flux quantum.
The combination
$\xi^{1/2} d^{3/2}$ in the denominator corresponds to an area enclosed between two typical diffusive trajectories connecting points ${\bm r}$ and ${\bm r}^\prime$. The condition $B \ll H_{\rm orb}$ means that the flux through this area is small in comparison with $\Phi_0$.

We see from Eq.~\eqref{eq:mag_field} that the field does not affect the matrix structure of $\langle {\cal G}_{\tau \tau^\prime} (\bm{r};\bm{r}^\prime | 0){\cal G}_{\rho^\prime \rho} (\bm{r}^\prime;\bm{r}| 0)\rangle$. Therefore, we expect that the condition $l_{\rm A} = l_{\rm N}$ remains intact even for  $l_{\rm A}$ and $l_{\rm N}$ modified by a field $B \gtrsim H_{\rm orb}$.

\section{Derivation of $\langle \delta \Theta_j^2\rangle$ for the forward scattering phase $\delta \Theta_j$}
In this section, we compute the variance of the forward scattering phase $\langle \delta \Theta_j^2\rangle$. Although its particular value is inconsequential for the results discussed in the main text, we present it here for completeness. The forward scattering phase $\delta \Theta_j$ accumulated by an electron across a short element can be found treating $H_{\rm prox}$ [Eq.~(2) of the main text] in the Born approximation. Using the resulting perturbative expression for $\delta \Theta_j$ in $\langle \delta \Theta_j^2\rangle$ and repeating the steps leading to Eq.~\eqref{eq:var3}, we obtain:
\begin{equation}
    \langle \delta \Theta_j^2\rangle = \,\frac{\pi^3 \nu_{\rm M} p_F^2 t^4}{v^2}\int_x^{x + \delta L}\hspace{-0.3cm} dx_1 dx_1^\prime \int_0^{+\infty}d\tau \,e^{-2\Delta \tau} \bigl[{\cal D}_{\uparrow\uparrow, \uparrow\uparrow}(x_1,y_j;x_1^\prime,y_j|\tau) + {\cal C}_{\uparrow\uparrow, \uparrow\uparrow}(x_1,y_j;x_1^\prime,y_j|\tau)\bigr].
\end{equation}
Here ${\cal C}_{\uparrow\uparrow, \uparrow\uparrow}$ is the normal-state Cooperon; in contrast to the calculations of Sec.~\ref{sec:lAlN}, its contribution cannot be neglected in the present derivation. Similarly to diffuson though, the Cooperon can be related to the diffusion kernel ${\cal D}(\bm{r};\bm{r}^\prime|\tau)$:
\begin{equation}\label{eq:cooperon}
{\cal C}_{\uparrow\uparrow, \uparrow\uparrow}(\bm{r};\bm{r}^\prime|\tau) = \exp\Bigl(- \frac{4\tau}{3\tau_{\rm so}}\Bigr) {\cal D}(\bm{r};\bm{r}^\prime|\tau).
\end{equation}
For an electrode of width $d \gg \xi$, it is possible to approximate the diffusion kernel by ${\cal D}(x_1,y_j;x_1^\prime,y_j|\tau) = 2\cdot e^{-(x_1 - x_1^\prime)^2 / 4D\tau} / (4\pi D \tau)^{3/2}$. Using this expression together with Eqs.~\eqref{eq:diff_kernel} and \eqref{eq:cooperon}, we find:
\begin{equation}\label{eq:fwd}
    \langle \delta \Theta_j^2\rangle = \,\frac{\pi^2 \nu_{\rm M} p_F^2 t^4}{2 v^2 D}\cdot \delta L \int \frac{d(\delta x)}{|\delta x|}\Bigl[\frac{1}{2}e^{-|\delta x|/\xi} + \frac{3}{2}e^{-|\delta x|/\xi^\star}\Bigr] \equiv \frac{\delta L}{l_{\rm F}},\quad\quad \frac{1}{l_{\rm F}} = \frac{4\pi g^2}{G_Q \sigma} \ln\Bigl[\frac{\xi^{1/4}(\xi^\star)^{3/4}}{l_{\rm mfp}}\Bigr].
\end{equation}
Parameter $\xi^\star$ here is defined in Eq.~\eqref{eq:diff_kernel}. The logarithmic factor originates from the diffusive character of electron motion in the metal \cite{us}. We cut-off the logarithmic divergence of the integral over $x$ at a distance $\sim l_{\rm mfp}$.

\section{Derivation of Eq.~(9) for the evolution of the $S$-matrix}
\begin{figure}[t!]
\begin{center}
\includegraphics[scale = 1]{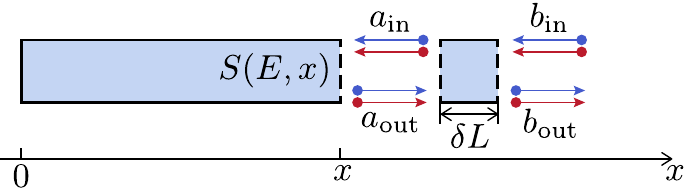}
\end{center}
\vspace{-0.1in}\caption{\small Schematics of the scattering problem for the particle-hole waves.  
}
\label{fig:1}
\end{figure}

Here we present details on how we obtain Eq.~(9) for the evolution of the $S$-matrix from the results of section ``Scattering off a short element''. 
We consider an addition of a short element of length $\delta L$ to a superconducting electrode of length $x$. Before the element was added, the $S$-matrix of the electrode was $S(E,x)$. After the element addition it changes to $S(E, x + \delta L)$. The two introduced $S$-matrices relate the amplitudes of outgoing and incoming particle-hole waves in the following way:
\begin{equation}\label{eq:S_old_and_new}
    a_{\rm out} = S(E, x)a_{\rm in}, \quad\quad b_{\rm out} = S(E, x + \delta L) b_{\rm in},
\end{equation}
where $a_{i} \equiv (a_{i, {\rm e}}, a_{i, {\rm h}})^T$, $b_{i} \equiv (b_{i, {\rm e}}, b_{i, {\rm h}})^T$ [see Fig.~\ref{fig:1}]. The change $\delta S(E,x) \equiv S(E, x + \delta L) - S(E,x)$ can be related to the scattering matrix of the added element $S_{\delta L}(E, x)$. The latter is a $4 \times 4$ matrix satisfying
\begin{equation}\label{eq:S_bridge}
    \begin{pmatrix}
    b_{\rm out}\\
    a_{\rm in}
    \end{pmatrix}
    =
    S_{\delta L} (E, x)
    \begin{pmatrix}
    a_{\rm out}\\
    b_{\rm in}
    \end{pmatrix}.
\end{equation}
In the main text, we found a subset of $S_{\delta L} (E, x)$ entries. 
The remaining entries can be obtained with the help of the particle-hole symmetry $S_{\delta L} (E , x) = \tau_x S_{\delta L}^\star(-E, x) \tau_x$ and unitarity condition (to start with, we disregard the quasiparticle loss due to vortices). 
Since the calculations in the main text were performed at the Fermi level, here we also generalize them to a finite energy. This is achieved by replacing the phase exponents in Eqs.~(4a) and (4b) by $e^{-i(k(E)x^\prime - k(-E)x)}$ and $e^{-ik(E)(x^\prime + x)}$, respectively, where $k(E) = k_\mu + E / v$. In this way, we find for a short element:
\begin{equation}\label{eq:S_bridge_entries}
    S_{\delta L}(E, x) \simeq 1  - i
    \begin{pmatrix}
    \delta {\cal R}(x) & \delta {\cal U}(x) e^{-2iEx/v}\\
    \delta {\cal U}^\dagger(x) e^{2iEx/v} & \delta {\cal L}(x)
    \end{pmatrix}
    .
\end{equation}
The $2\times 2$ matrices $\delta {\cal R}(x)$ and $\delta {\cal L}(x)$ contain the forward scattering amplitudes; matrix $\delta {\cal U}(x)$ describes CAR and EC processes:
\begin{equation}
    \delta {\cal R}(x) =
    \begin{pmatrix}
    - \delta \Theta_{\rm R}(x) & 0\\
    0  & \delta \Theta_{\rm R}(x)
    \end{pmatrix},
    \quad\quad
    \delta {\cal L}(x) =
    \begin{pmatrix}
    - \delta \Theta_{\rm L}(x) & 0\\
    0  & \delta \Theta_{\rm L}(x)
    \end{pmatrix},
    \quad\quad
    \delta {\cal U}(x) =
    \begin{pmatrix}
    \delta A_{\rm N}(x) & \delta A_{\rm A}(x)\\
    -\delta A_{\rm A}^\star(x) & -\delta A_{\rm N}^\star(x)
    \end{pmatrix}.
\end{equation}
Random variables $\delta A_{\rm A/N}(x)$ satisfy Eq.~(5) of the main text, while $\delta \Theta_{\rm R/L}(x)$ satisfy a respective relation \eqref{eq:fwd}.

Next, we use Eqs.~\eqref{eq:S_old_and_new} and \eqref{eq:S_bridge} to express $S(E, x + \delta L)$ in terms of $S(E,x)$ and $S_{\delta L} (E, x)$. With the help of Eq.~\eqref{eq:S_bridge_entries} we find to the order $\sqrt{\delta L}$:
\begin{equation}\label{eq:dS}
    \delta S(E,x) \equiv S(E, x + \delta L) - S(E, x) = {\delta {\cal U}(x)}e^{-2iEx/v} + S(E,x) {\delta {\cal U}^\dagger (x)} S(E, x) e^{2iEx/v} + S(E,x){\delta {\cal L}(x)} + {\delta {\cal R}(x)}S(E,x).
\end{equation}
It is further convenient to reparameterize the $S$-matrix by separating out the energy-dependent common phase, $S(E, x) \rightarrow S(E, x) e^{-2iEx/v}$. Upon doing that, and promoting a finite difference equation \eqref{eq:dS} to a differential one, we obtain Eq.~(9) of the main text, with $\Gamma = 0$. 

Finally, we account for the presence of extra scattering channels associated with the dense spectrum of levels in the vortex cores. 
We assume that the tunneled into a vortex core quasiparticle does not return coherently to the edge; this is justified if the escape time of a quasiparticle from the core exceeds its inelastic scattering time within the core.
We describe the induced by the vortices quasiparticle loss phenomenologically, by adding a term $-2\Gamma \delta L / v$ to the right hand side of Eq.~\eqref{eq:S_bridge_entries}. It is this term that produces a contribution $-2i \Gamma S / v$ to the right hand side of Eq.~(9) of the main text.

\section{Derivation of Eq.~(13) for the dynamics of variables $w_i$}
In this section, we present details of the derivation of Eq.~(13) for the dynamics of variables $w_1$ and $w_2$. 
To obtain a system of equations that would focus exclusively on these two variables, we need to account for correlations that exist between the amplitudes $\eta_{\rm A / N}$ and the remaining variables $\alpha$ and $\phi$ . 
These correlations lead to an extra ``ponderomotive'' potential for the $\vec{w}$-particle [see second and third terms in Eq.~(14) of the main text]. To account for the correlations, we first solve Eqs.~(12c) and (12d) formally in a short interval $[x_-, x]$ with $x_- = x - \delta L < x$:
\begin{subequations}
\begin{align}
    \alpha(x) &= \alpha(x_-) + \int_{x_-}^x dx^\prime q_{12}(x^\prime)\bigl[\eta_{{\rm N}x}(x^\prime) \cos \alpha(x^\prime) + \eta_{{\rm N}y}(x^\prime) \sin \alpha(x^\prime)\bigr] + \int_{x_-}^x dx^\prime [\vartheta_{\rm R}(x^\prime) + \vartheta_{\rm L}(x^\prime)],\label{eq:alpha_scary}\\
    \phi(x) &= \phi(x_-) - {\rm s}_{12}(x^\prime) \int_{x_-}^x dx^\prime q_{21}(x^\prime)\bigl[\eta_{{\rm A}x}(x^\prime) \cos \phi(x^\prime) + \eta_{{\rm A}y}(x^\prime) \sin \phi(x^\prime)\bigr] + \int_{x_-}^x dx^\prime [\vartheta_{\rm R}(x^\prime) - \vartheta_{\rm L}(x^\prime)].\label{eq:phi_scary}
\end{align}
\end{subequations}
Here, to make the notations concise, we abbreviated ${\rm s}_{12}(x^\prime) \equiv {\rm sign}(w_1(x^\prime) - w_2(x^\prime))$, $q_{12}(x^\prime) \equiv q(w_1(x^\prime), w_2(x^\prime))$, $q_{21}(x^\prime) \equiv  q(w_2(x^\prime), w_1(x^\prime))$. Function 
\begin{equation}\label{eq:q-function}
    q(w_1, w_2) = \tanh \frac{w_1 + w_2}{2}\Theta(w_1 - w_2) + \coth \frac{w_1 - w_2}{2} \Theta(w_2 - w_1)
\end{equation}
was introduced in the main text after Eq.~(12). 
It can be verified easily that the forward scattering phases $\vartheta_{\rm R}$ and $\vartheta_{\rm L}$ are inconsequential for the present derivation.
Therefore, we suppress them in all of the following expressions.
In the relevant limit $\delta L \rightarrow 0$, it is possible to approximate $w_i(x^\prime) \approx w_i(x_-)$, $\alpha(x^\prime) \approx \alpha (x_-)$, and $\phi(x^\prime) \approx \phi (x_-)$ in Eqs.~\eqref{eq:alpha_scary} and \eqref{eq:phi_scary}. Then, these equations simplify to
\begin{subequations}
\begin{align}
    \alpha(x) &\approx \alpha(x_-) +  q_{12}(x_-)\bigl[\cos \alpha(x_-) \int_{x_-}^x dx^\prime\eta_{{\rm N}x}(x^\prime) + \sin \alpha(x_-) \int_{x_-}^x dx^\prime\eta_{{\rm N}y}(x^\prime)\bigr],\\
    \phi(x) &\approx \phi(x_-) -  {\rm s}_{12}(x_-)q_{21}(x_-)\bigl[\cos \phi(x_-)\int_{x_-}^x dx^\prime \eta_{{\rm A}x}(x^\prime)  -  \sin \phi(x_-)\int_{x_-}^x dx^\prime \eta_{{\rm A}y}(x^\prime)\bigr].
\end{align}
\end{subequations}
We now substitute these expressions into Eqs.~(12a) and (12b) of the main text. Expanding sines and cosines in Taylor series, we obtain to the second order in $\eta_{\rm A/ N}$:
\begin{align}
    \frac{dw_1(x)}{dx} = &\frac{2E}{v} \cosh{w_1(x)} + \tilde{\eta}_1(x)+ q_{12}(x_-)\Bigl[ \cos^2 \alpha(x_-) \int_{x_-}^{x} dx^\prime \eta_{{\rm N}x}(x)\eta_{{\rm N}x}(x^\prime) + \sin^2 \alpha(x_-) \int_{x_-}^{x} dx^\prime \eta_{{\rm N}y}(x)\eta_{{\rm N}y}(x^\prime)\Bigr]\notag\\
    &-\,{\rm s}_{12}(x_-) q_{21}(x_-)\Bigl[ \cos^2 \phi(x_-) \int_{x_-}^{x} dx^\prime \eta_{{\rm A}x}(x)\eta_{{\rm A}x}(x^\prime) + \sin^2 \phi(x_-) \int_{x_-}^{x} dx^\prime \eta_{{\rm A}y}(x)\eta_{{\rm A}y}(x^\prime)\Bigr] + \dots,\label{eq:w1}
\end{align}
and similarly
\begin{align}
    \frac{dw_2(x)}{dx} = &\frac{2E}{v} \cosh{w_2(x)} + \tilde{\eta}_2(x)+ {\rm s}_{12}(x_-) q_{12}(x_-)\Bigl[ \cos^2 \alpha(x_-) \int_{x_-}^{x} dx^\prime \eta_{{\rm N}x}(x)\eta_{{\rm N}x}(x^\prime) + \sin^2 \alpha(x_-) \int_{x_-}^{x} dx^\prime \eta_{{\rm N}y}(x)\eta_{{\rm N}y}(x^\prime)\Bigr]\notag\\
    &+ \, q_{21}(x_-)\Bigl[ \cos^2 \phi(x_-) \int_{x_-}^{x} dx^\prime \eta_{{\rm A}x}(x)\eta_{{\rm A}x}(x^\prime) + \sin^2 \phi(x_-) \int_{x_-}^{x} dx^\prime \eta_{{\rm A}y}(x)\eta_{{\rm A}y}(x^\prime)\Bigr] + \dots \label{eq:w2}
\end{align}
Here $\dots$ denotes contributions which contain products $\eta_{{\rm N}, i} \eta_{{\rm N},j}$ and  $\eta_{{\rm A}, i} \eta_{{\rm A},j}$ with $i\neq j$. These contributions vanish in the limit $\delta L \rightarrow 0$, as follows from Eq.~(8) of the main text. Terms $\tilde{\eta}_{1/2}(x)$ in Eqs.~\eqref{eq:w1} and \eqref{eq:w2} are given by
\begin{subequations}
\begin{align}
    \tilde{\eta}_{1}(x) &= \eta_{{\rm N}x}(x)\sin \alpha(x_-) - \eta_{{\rm N}y}(x)\cos \alpha(x_-) + \eta_{{\rm A}x} (x) \sin \phi(x_-)
    - \eta_{{\rm A}y}(x) \cos \phi(x_-),\\
    \tilde{\eta}_{2}(x) &= {\rm s}_{12}(x)\bigl[\eta_{{\rm N}x}(x)\sin \alpha(x_-) - \eta_{{\rm N}y}(x)\cos \alpha(x_-) - \eta_{{\rm A}x} (x) \sin \phi(x_-)
    + \eta_{{\rm A}y}(x) \cos \phi(x_-)\bigr].
\end{align}
\end{subequations}
Importantly,  because $x_- < x$, $\eta_{\rm N}(x)$ and $\eta_{\rm A}(x)$ here are statistically independent of $\alpha(x_-)$ and $\phi(x_-)$. Then, we can readily find the correlators of $\tilde{\eta}_i(x)$ with the help of Eq.~(8) of the main text. Focusing first on the critical point, $l_{\rm A} = l_{\rm N} = 2l_0$, we obtain:
\begin{equation}\label{eq:corr_func_tilde}
    \langle \tilde{\eta}_i(x_1) \tilde{\eta}_j(x_2) \rangle = \frac{\delta (x_1 - x_2) \delta_{ij}}{l_0}.
\end{equation}
To further simplify Eqs.~\eqref{eq:w1} and \eqref{eq:w2}, let us consider one of the integrals entering these expressions, e.g., $\int_{x_-}^{x} dx^\prime \eta_{{\rm N}x}(x)\eta_{{\rm N}x}(x^\prime)$. It is convenient to split this integral into its disorder-averaged value and a fluctuation component,
$\int_{x_-}^{x} dx^\prime \langle\eta_{{\rm N}x}(x)\eta_{{\rm N}x}(x^\prime)\rangle + \delta [\int_{x_-}^{x} dx^\prime \eta_{{\rm N}x}(x)\eta_{{\rm N}x}(x^\prime)]$. The fluctuation component vanishes in the limit $x_- \rightarrow x$. On the other hand, the average part does not. Indeed, using Eq.~(8) of the main text, we obtain  $\int_{x_-}^{x} dx^\prime \langle \eta_{{\rm N}x}(x)\eta_{{\rm N}x}(x^\prime)\rangle  = 1/(2 l_0)$ [we tacitly assume here that $\delta L = x - x_-$ still exceeds the ``microscopic'' scale $\sim \sqrt{\xi d}$ below which the fields $\eta_{\rm A/N}(x)$ can no longer be treated as delta-correlated ones]. 
We can compute the other integrals in Eqs.~\eqref{eq:w1} and \eqref{eq:w2} in the same way, $\int_{x_-}^x dx^\prime  \eta_i(x)\eta_i(x^\prime) = 1/(2l_0)$ ($i \in \{ {\rm N}x, {\rm N}y, {\rm A}x, {\rm A}y\}$). Thus, at the critical point, we arrive at
\begin{align}
    \frac{dw_i(x)}{dx} = \frac{2E}{v} \cosh{w_i(x)} + \frac{1}{l_0} \frac{\sigma_i \cosh w_i(x)}{\sinh w_1(x) - \sinh w_2(x)} + \tilde{\eta}_i(x),
\end{align}
where we used Eq.~\eqref{eq:q-function} and defined $\sigma_{1/2} = \pm 1$. Introducing the effective potential $U(w_1, w_2)$ [see Eq.~(14) of the main text with $\lambda$ set to zero], we find Eq.~(13).

A deviation $\lambda$ from the critical point changes both the correlation function \eqref{eq:corr_func_tilde} and the effective potential $U(w_1, w_2)$ [we remind that $\lambda$ parameterizes the difference in $l_{\rm A}$ and $l_{\rm N}$, i.e., $l_{\rm A} = 2l_0(1+\lambda)$ and $l_{\rm N} = 2l_0(1-\lambda)$]. However, for $|\lambda| \ll 1$, the change in the correlation function has a negligible influence on the character of the $\vec{w}$\,-particle dynamics, and we disregard it in the main text. At the same time, the change in the effective potential has important ramifications for the low-energy conductance and density of states. The correction to $U(w_1, w_2)$ due to $|\lambda| \ll 1$ reads
\begin{equation}
    U_\lambda(w_1, w_2) = -{\rm sign}(w_1-w_2)\frac{\lambda}{l_0}\Bigl[\ln\cosh \frac{w_1 + w_2}{2}-\ln \sinh \frac{|w_1 - w_2|}{2} \Bigr],
\end{equation}
as follows directly from Eqs.~\eqref{eq:w1}, \eqref{eq:w2}, and Eq.~(8) of the main text. This expression can be simplified significantly in the low-energy regime. Keeping in mind that the motion of the $\vec{w}$-patricle in this regime is confined to the two trenches of length $\sim \ln (v / El_0) \gg 1$ [see Fig.~2 of the main text], we can approximate $U_\lambda(w_1, w_2) \approx \lambda (w_1 - w_2) / l_0$ up to a constant offset (we note that the offset is different for $w_1 > w_2$ and $w_1 < w_2$). The latter expression constitutes the third term of Eq.~(14) of the main text.

\section{Derivation of the distribution function $P(G)$ at arbitrary $\Gamma$}
In this section, we generalize Eq.~(23) for the conductance distribution function $P(G)$ to arbitrary values of the quasiparticle loss rate $\Gamma$. We start with Eq.~(8) of the main text, in which we take $E = 0$. A convenient parameterization of the $S$-matrix at $E = 0$ and non-zero $\Gamma$ is:
\begin{equation}
    S = -\frac{1}{2}
    \begin{pmatrix}
    \Bigl(\tanh \frac{m_1}{2} + \tanh \frac{m_2}{2}\Bigr) e^{i\alpha} & \Bigl(\tanh \frac{m_2}{2} - \tanh \frac{m_1}{2}\Bigr)e^{i\phi}\\
    \Bigl(\tanh \frac{m_2}{2} - \tanh \frac{m_1}{2}\Bigr) e^{-i\phi} & \Bigl(\tanh \frac{m_2}{2} + \tanh \frac{m_1}{2}\Bigr)e^{-i\alpha}
    \end{pmatrix}.
\end{equation}
Substituting this representation in Eq.~(8), we obtain a system of equations for variables $m_1, m_2, \alpha, \phi$:
\begin{subequations}\label{eq:evol_with_loss}
\begin{align}
    \frac{dm_1}{dx} &= - \frac{2\Gamma}{v} \sinh m_1 + \eta_{{\rm N} x} \sin \alpha - \eta_{{\rm N} y} \cos \alpha + \eta_{{\rm A} x} \sin \phi - \eta_{{\rm A} y} \cos \phi, \label{eq:m_1_evol_g}\\ 
    \frac{dm_2}{dx} &= -\frac{2\Gamma}{v} \sinh m_2 + \eta_{{\rm N} x} \sin \alpha - \eta_{{\rm N} y} \cos \alpha - \eta_{{\rm A} x} \sin \phi + \eta_{{\rm A} y} \cos \phi, \label{eq:m_2_evol_g}\\
    \frac{d\alpha}{dx} &= \vartheta_{\rm R} + \vartheta_{\rm L} +  \coth \frac{m_1+m_2}{2}\bigl(\eta_{{\rm N}x} \cos \alpha + \eta_{{\rm N}y} \sin \alpha\bigr), \label{eq:alpha_evol_g}\\
    \frac{d\phi}{dx} &= \vartheta_{\rm R} - \vartheta_{\rm L} +  \coth\frac{m_1-m_2}{2}\bigl(\eta_{{\rm A}x} \cos \phi + \eta_{{\rm A}y} \sin \phi\bigr). \label{eq:phi_evol_g}
\end{align}
\end{subequations}
The initial conditions are $m_1(x = 0) = m_2(x = 0) = -\infty$, $\alpha (0) = 0$, and $\phi(0) = 0$. The conductance can be expressed in terms of the values of $m_1$ and $m_2$ at $x = L$:
\begin{equation}\label{eq:cond_vortex}
    G = G_Q \tanh \frac{m_1(L)}{2} \tanh  \frac{m_2(L)}{2}.
\end{equation}
Finding $P(G)$ thus requires the knowledge of how variables $m_1(L) $ and $m_2(L)$ are distributed. To determine this, we derive a Fokker-Planck equation for the evolution of the distribution function $P(m_1, m_2|x) \equiv \int \frac{d\alpha}{2\pi} \frac{d\phi}{2\pi} P(m_1, m_2, \alpha, \phi|x)$. Applying a standard machinery \cite{vankampen} to Eq.~\eqref{eq:evol_with_loss}, we obtain
\begin{equation}\label{eq:fokker}
    \frac{\partial P}{\partial x} + \sum_{i = 1,2} \frac{\partial j_i}{\partial m_i} = 0,\quad\quad j_i(m_1,m_2) = - \frac{\partial U}{\partial m_i}P - \frac{1}{l_0} \frac{\partial P}{\partial m_i},
\end{equation}
where $j_i(m_1,m_2)$ has a meaning of the probability current, and
\begin{equation}\label{eq:conf_potential}
    U(m_1,m_2) = \frac{2\Gamma}{v}(\cosh m_1 + \cosh m_2) - \frac{1}{l_0}\ln |\cosh m_1 - \cosh m_2|. 
\end{equation}
The Fokker-Planck equation has a form similar to a diffusion equation for a Brownian particle moving in an external field with a potential $U(m_1, m_2)$. For sufficiently large sample size $L$ the distribution function reaches a steady state $P_{\rm st}(m_1, m_2)$. Because the potential is confining [cf.~Eq.~\eqref{eq:conf_potential}], we can find the steady state by demanding that the probability  current  $j_i(m_1, m_2)$ vanishes.
The resulting steady state has the Gibbs form, $P_{\rm st} \propto \exp[-l_0\,U(m_1, m_2)]$.

We can now find the conductance distribution function as $P(G) = \langle \delta (G - G_Q \tanh(m_1/2)\tanh(m_2/2))\rangle$, where the average is performed over $P_{\rm st}(m_1, m_2)$. After a straightforward calculation, this leads to the following result for $P(G)$:
\begin{equation}\label{eq:PG_vortices}
    P(G) = \frac{1}{-{\rm li}(e^{-4\Gamma l_0 / v})} \frac{G_Q}{G_Q^2 - G^2}\exp\Bigl(-\frac{4\Gamma l_0}{v} \frac{G_Q+|G|}{G_Q - |G|}\Bigr)
\end{equation}
Here ${\rm li}(x) = \int_0^x d\rho/\ln \rho$ is logarithmic integral function (note that ${\rm li}(x) < 0$ for $0<x<1$). The expression above reduces to Eq.~(23) of the main text in the regime of strong quasiparticle absorption, $\Gamma l_0 / v \gg 1$. However, in contrast to Eq.~(23), Eq.~\eqref{eq:PG_vortices} is applicable at arbitrary $\Gamma l_0 / v$. We used the latter equation to produce curves in Fig.~4.

\section{Calculation of $\langle \delta A_{\rm A}(x; k_\mu) \delta A^\star_{\rm A}(x; k_\mu+\delta k_\mu)\rangle$}
Here we compute the correlator $\langle \delta A_{\rm A}(x; k_\mu) \delta A^\star_{\rm A}(x; k_\mu+\delta k_\mu)\rangle$ of the CAR amplitudes at different values of the edge states Fermi momentum $k_\mu$. We use this correlator to obtain the first term in Eq.~(24) of the main text. The calculation is largely similar to the one presented in Sec.~\ref{sec:lAlN}. A counterpart of Eq.~\eqref{eq:var1} is
\begin{align}
        \langle \delta A_{\rm A}(x; k_\mu) \delta A_{\rm A}^\star(x; k_\mu + \delta k_\mu) \rangle = \frac{t^4}{v^2}\int^{x + \delta L}_{x}\hspace{-0.3cm}
        dx_1 dx_1^\prime&  dx_2 dx_2^\prime\,
        e^{ik_\mu(x_1  - x_1^\prime)} e^{- i(k_\mu + \delta k_\mu) (x_2 - x_2^\prime)}\notag\\
        &\times\partial^2_{y_1,y_1^\prime}\partial^2_{y_2, y_2^\prime}\Lngl {\cal G}^{\downarrow \uparrow}_{\rm he}(\bm{r}_1 ; \bm{r}_1^\prime|E = 0)\cdot {\cal G}^{\uparrow \downarrow}_{\rm eh}(\bm{r}_2^\prime ; \bm{r}_2|E = 0)\Rngl\Bigr|^{y_1^\prime, y_2^\prime = d}_{y_1, y_2 = 0}.
\end{align}
The only difference between this expression and Eq.~\eqref{eq:var1} is in the exponential factors, which depend on $k_\mu$. 
Similarly to what we did it in Sec.~\ref{sec:lAlN}, we relate the average of the product of the Green's functions to the normal-state diffuson. By repeating the steps leading to Eq.~\eqref{eq:vars_final}, we arrive to the following expression for $d \gg \xi$:
\begin{equation}
    \langle \delta A_{\rm A}(x; k_\mu) \delta A_{\rm A}^\star(x; k_\mu + \delta k_\mu) \rangle = \frac{\pi^2\nu_{\rm M} p_F^2 t^4}{2 v^2 D }\cdot \delta L \int \frac{d(\delta x)}{\sqrt{d^2 + (\delta x)^2}} e^{-i\delta k_\mu \delta x}\Bigl[e^{-\sqrt{d^2 + (\delta x)^2}/\xi} -  e^{-\sqrt{d^2 + (\delta x)^2}/\xi^\star} \Bigr]
\end{equation}
(see Eq.~\eqref{eq:vars_final} for the definition of $\xi^\star$). 
In the limit of strong spin-orbit interaction, which we shall focus on, the second term in the square brackets can be neglected. Disposing with it and also expanding the square roots in Taylor series, we obtain
\begin{equation}\label{eq:A_A_k_mu}
    \langle \delta A_{\rm A}(x; k_\mu) \delta A_{\rm A}^\star(x; k_\mu + \delta k_\mu) \rangle = \frac{\pi^2\nu_{\rm M} p_F^2 t^4}{2 v^2 D }\cdot \delta L \int \frac{dx}{d} e^{-i\delta k_\mu x} e^{-d/\xi - x^2 /(2\xi d)} \equiv \frac{\delta L}{2l_0} \exp\bigl( - \delta k_\mu^2 \xi d\bigr).
\end{equation}
Exponent $\exp(-d/\xi - (\delta x)^2/(2\xi d))$ here 
stems from
the probability of the quasiparticle tunneling 
with a displacement
$\delta x$ in the lateral direction.
The characteristic lateral displacement $\delta x \sim \sqrt{\xi d}$ sets the scale $\sim 1/\sqrt{\xi d}$ for the variation of $\delta A_{\rm A}(x; k_\mu)$ with $k_\mu$. By promoting $\delta A_{\rm A}(x; k_\mu) \rightarrow \eta_{\rm A}(x; k_\mu)\cdot \sqrt{\delta L} / 2$, and using Eq.~\eqref{eq:A_A_k_mu} together with Eq.~(22) of the main text, we obtain the first term in the expression for ${\cal C}_{n}(\delta n)$, see Eq.~(24).

\section{Demonstration of ${\cal C}_n(\delta n)$ saturation for $\Gamma \rightarrow 0$\label{sec:saturation}}
In this section, we show that the correlation function ${\cal C}_n(\delta n)$ saturates at an energy-independent value $\simeq 0.31$ for $n_{\rm c, N} \ll \delta n \ll n_{\rm c, A}$ and $\Gamma = 0$. Here $n_{\rm c, N}$ and $n_{\rm c, A}$ are the two scales of the conductance variation with electron density, which are associated with variations of EC and CAR amplitudes, respectively.

To start with, we note that system of equations (12) [see the main text] can be significantly simplified in the low-energy limit, $E \ll v / l_0$. 
Indeed, in this limit, the motion of the $\vec{w}$\,-``particle'' is essentially one-dimensional: it is occurs along two straight, equipotential trenches of length $2\ln (v / E l_0) \gg 1$. 
This makes it convenient to introduce a \textit{single} variable $w$ parameterizing the position of the particle along the trenches. 
We define the variable $w$ on an interval $[0, 4\ln(v / E l_0)]$.
The first half of the interval
corresponds to the vertical trench of Fig.~(2), while its second half
corresponds to the horizontal trench. 
The evolution of the $w$-variable is cyclic.
As $w$ reaches the point $4\ln (v/E l_0)$, it resets to $w = 0$; in terms of Fig.~(2), this corresponds to a transition from the horizontal trench to the vertical one.
Finally, the points $w = 2\ln(v / E l_0), 4\ln(v / E l_0)$ act as the ``cliffs'': the particle can freely pass these points from left to right, but never in the opposite direction.
Another simplification of the low-energy limit comes from the behavior of functions $q(w_1, w_2)$ and $q(w_2, w_1)$ in Eq.~(12). 
These functions can be approximated by $\pm 1$ everywhere along the trenches, except for narrow intervals of relative width $\sim 1 / \ln(v / E l_0)$ near the trenches end-points. The relative width vanishes for $E \rightarrow 0$, so we can neglect the influence of these intervals on the $S$-matrix dynamics.
Applying the above approximations to Eq.~(12) of the main text, we obtain a system of coupled equations for three variables, i.e., $w,\alpha$, and $\phi$:
\begin{subequations}\label{eq:w_evol_system}
\begin{align}\label{eq:w_evol}
    \frac{dw}{dx} &= \sigma(w) (\eta_{{\rm N} x} \sin \alpha - \eta_{{\rm N} y} \cos \alpha) +
    \eta_{{\rm A} x} \sin \phi - \eta_{{\rm A} y} \cos \phi,\\
    \frac{d\alpha}{dx} &= \vartheta_{\rm R} + \vartheta_{\rm L} - 
    (\eta_{{\rm N}x} \cos \alpha + \eta_{{\rm N}y} \sin \alpha\bigr), \label{eq:alpha_evol_x}\\
    \frac{d\phi}{dx} &= \vartheta_{\rm R} - \vartheta_{\rm L} +  \sigma(w)\bigl(\eta_{{\rm A}x} \cos \phi + \eta_{{\rm A}y} \sin \phi\bigr), \label{eq:phi_evol_x}
\end{align}
\end{subequations}
where we introduced a sign-function $\sigma(w) = {\rm sign}(2\ln(v/E l_0) - w)$.

At this point, it is important to comment on a nuance in the employed $S$-matrix parameterization [see Eq.~(11) of the main text]. 
While the evolution of $S(E,x)$ with $x$ is obviously continuous [see Eq.~(8)], the evolution of variables $\alpha$ and $\phi$ is not.
Indeed, a direct inspection of Eq.~(11) shows that variables $\alpha$ and $\phi$ have to slip by $\pi$ every time $\vec{w}$ transitions from the vertical trench to the horizontal one; in fact, it is the continuity of the $S$-matrix that necessitates these phase slips.
To account for the phase slips in \eqref{eq:w_evol_system}, it is convenient to introduce auxiliary variables $\alpha_0$ and $\phi_0$, which---in contrast to $\alpha$ and $\phi$---would change continuously with $x$ and would be independent of $w$. We define $\alpha_0$ and $\phi_0$ by the following equations:
\begin{subequations}\label{eq:alpha_0_and_phi_0}
\begin{align}
    \frac{d\alpha_0}{dx} &= \vartheta_{\rm R} + \vartheta_{\rm L} - \bigl(\eta_{{\rm N}x} \cos \alpha_0 + \eta_{{\rm N}y} \sin \alpha_0\bigr), \label{eq:alpha_evol_t_simpl}\\
    \frac{d\phi_0}{dx} &= \vartheta_{\rm R} - \vartheta_{\rm L} + \bigl(\eta_{{\rm A}x} \cos \phi_0 + \eta_{{\rm A}y} \sin \phi_0\bigr). \label{eq:phi_evol_t_simple}
\end{align}
\end{subequations}
Using Eqs.~\eqref{eq:alpha_evol_x}, \eqref{eq:phi_evol_x}, and \eqref{eq:alpha_0_and_phi_0}, it is easy to show that variables $\alpha$ and $\phi$ are almost always close to $\alpha_0$ and $\phi_0$, respectively, up to an integer multiple of $\pi$:
\begin{equation}\label{eq:connection}
    \alpha(x) = \alpha_0(x) + \pi n_\alpha,\quad\quad \phi(x) = \phi_0(x) + \pi n_\phi.
\end{equation}
Here, the multiple $n_\alpha$ is an even number, while the parity of $n_\phi$ depends on $\sigma(w)$: $n_\phi$ is odd if $\sigma(w) = -1$ and even if $\sigma(w) = +1$. 
Connection \eqref{eq:connection} breaks for transient intervals of ``time'' $\Delta x \sim l_0$ only, which occur whenever $w$ passes through points $w = 2\ln (v / E l_0)$ or $4\ln (v / E l_0)$.
The fraction of ``time'' spanned by these intervals can be estimated as $\sim 1 / \ln^2(v/El_0)$ and is thus negligible in the in the low-energy limit.
Then, using Eq.~\eqref{eq:connection} in Eq.~\eqref{eq:w_evol}, we arrive at a concise equation for the evolution of $w$ with $x$:
\begin{align}\label{eq:w_evol_final}
    \frac{dw}{dx} = \sigma(w) (\eta_{\rm N 0} + \eta_{\rm A 0}),\quad\quad  \eta_{\rm N 0} \equiv \eta_{{\rm N} x} \sin \alpha_0 - \eta_{{\rm N} y} \cos \alpha_0,\quad\quad \eta_{\rm A 0} \equiv \eta_{{\rm A} x} \sin \phi_0 - \eta_{{\rm A} y} \cos \phi_0.
\end{align}
The fields $\eta_{\rm N 0}(x)$ and $\eta_{\rm A 0}(x)$ here are random and uncorrelated with each other. Using Eq.~(8) of the main text, we find that their correlators are given by
$\langle \eta_{\rm N 0}(x) \eta_{\rm N 0}(x^\prime)\rangle = \delta (x - x^\prime) / l_0$ and $\langle \eta_{\rm A 0}(x) \eta_{\rm A 0}(x^\prime)\rangle = \delta (x - x^\prime) / l_0$.

Let us now apply Eq.~\eqref{eq:w_evol_final} to demonstrate the saturation of the conductance correlation function ${\cal C}_{n}(\delta n)$ for $n_{\rm c, N} \ll \delta n \ll n_{\rm c, A}$. 
The conductance is related to the value of the $w$-variable at the end of its evolution (i.e., at $x = L$), $G = G_Q\,{\rm sign}(|2 - w(L) / \ln (v / E l_0)| - 1)$ [the latter expression follows from Eq.~(15) of the main text in the low-energy limit, also see Fig.~2].
Consequently, to find ${\cal C}_{n}(\delta n)$, we need to compare the results of the $w$-variable evolution at two values of electron density $n$ and $n + \delta n$.
The respective evolution equations read:
\begin{subequations}\label{eq:evol_diff_dens}
\begin{align}
    \frac{dw_n}{dx} &= \sigma (w_n) (\eta_{{\rm N 0,}n} + \eta_{\rm A 0}),\label{eq:evol_den_1}\\
    \frac{dw_{n+\delta n}}{dx} &= \sigma (w_{n+\delta n}) (\eta_{{\rm N 0,} n + \delta n} + \eta_{\rm A 0})\label{eq:evol_den_2}.
\end{align}
\end{subequations}
Because the considered density difference satisfies $\delta n \ll n_{\rm c, A}$, the amplitudes of CAR are the same for $w_n$ and $w_{n + \delta n}$. On the contrary, since $\delta n \gg n_{\rm c, N}$, one can treat the amplitudes of EC processes, $\eta_{{\rm N 0,}n}$ and $\eta_{{\rm N 0,} n + \delta n}$, as being statistically independent of each other. 

To characterize the correlations between $w_n(L)$ and $w_{n+\delta n}(L)$, we derive \cite{vankampen} a Fokker-Planck equation for the joint distribution function  $P(w_n, w_{n+\delta n}|x)$. Using Eqs.~\eqref{eq:evol_den_1} and \eqref{eq:evol_den_2}, we obtain 
\begin{equation}\label{eq:Pw}
    \frac{\partial P}{\partial x} = \underset{\text{EC}}{\underbrace{\frac{1}{l_0}\Bigl[\frac{\partial^2}{\partial w_{n}^2} +  \frac{\partial^2}{\partial w_{n+\delta n}^2}\Bigr] P}}
    +
    \underset{\text{CAR}}{\underbrace{\frac{1}{l_0}\Bigl[\frac{\partial^2}{\partial w_{n}^2} +  \frac{\partial^2}{\partial w_{n+\delta n}^2} + 2\sigma(w_{n})\sigma(w_{n+\delta n}) \frac{\partial^2}{\partial w_{n+\delta n}\partial w_{n+\delta n}}\Bigr] P}}.
\end{equation}
This equation is supplemented by the periodic boundary conditions with respect to both $w_n$ and $w_{n + \delta n}$ in the interval $[0, 4 \ln (v / E l_0)]$. Besides, we impose that points $w = 2\ln (v / E l_0)$ and $4 \ln (v / E l_0)$ act as the ``cliffs'', which means that the $w$-variable can pass through these points from left to right only (the precise mathematical formulation of this condition is not important for the present discussion).
Equation \eqref{eq:Pw} has a set of notable features.
Firstly, the variables $w_n$ and $w_{n+\delta n}$ in it do not separate. This reflects the fact that the CAR amplitudes are the same for $w_n$ and $w_{n + \delta n}$, cf.~Eq.~\eqref{eq:evol_diff_dens}; it is the ultimate reason for the saturation of ${\cal C}_n(\delta n)$ at a non-zero value.
Secondly, Eq.~\eqref{eq:Pw} admits a rescaling of variables of the form $w_i = \tilde{w}_i / \ln (v / E l_0)$ and $x = \tilde{x}\,l(E)$, where $l(E) = l_0 \ln^2 (v / E l_0)$ is the correlation radius at energy $E$. This rescaling makes the domain in which the $w$-variables are defined independent of $E$, so it will prove useful in demonstrating the energy independence of ${\cal C}_n(\delta n)$. In terms of the rescaled variables, Eq.~\eqref{eq:Pw} acquires the form:
\begin{equation}\label{eq:Pw_nondim}
    \frac{\partial \tilde{P}}{\partial \tilde{x}} = \Bigl[\frac{\partial^2}{\partial \tilde{w}_{n}^2} +  \frac{\partial^2}{\partial \tilde{w}_{n+\delta n}^2}\Bigr] \tilde{P}
    +
    \Bigl[\frac{\partial^2}{\partial \tilde{w}_{n}^2} +  \frac{\partial^2}{\partial \tilde{w}_{n+\delta n}^2} + 2\tilde{\sigma}(w_{n})\tilde{\sigma}(\tilde{w}_{n+\delta n}) \frac{\partial^2}{\partial \tilde{w}_{n+\delta n}\partial \tilde{w}_{n+\delta n}}\Bigr] \tilde{P}
\end{equation}
with $\tilde{\sigma}(\tilde{w}) = {\rm sign}(2 - \tilde{w})$ and $\tilde{P}(\tilde{w}_n, \tilde{w}_{n+\delta n}|\tilde{x}) \equiv P(w_n, w_{n+\delta n}|x) / \ln^2(v/E l_0)$,
and boundary conditions independent of $E$. 

The correlation function can be represented in terms of the rescaled variables as \begin{equation}\label{eq:corr_den}
    {\cal C}_n(\delta n) = \int_0^4 d\tilde{w}_n d\tilde{w}_{n+\delta n}\, \tilde{P}(\tilde{w}_n, \tilde{w}_{n+\delta n}| L / l(E))\,{\rm sign}(|2 - \tilde{w}_n| - 1)\,{\rm sign}(|2 - \tilde{w}_{n+\delta n}| - 1).
\end{equation}
Let us assume that the system size $L$ exceeds $l(E)$. In this case, the distribution function reaches a steady state $\tilde{P}_{\rm st}(\tilde{w}_n, \tilde{w}_{n+\delta n})$. From  Eq.~\eqref{eq:Pw_nondim}, it follows immediately  that $\tilde{P}_{\rm st}(\tilde{w}_n, \tilde{w}_{n+\delta n})$ is independent of $E$. Hence, the same holds for the right hand side of Eq.~\eqref{eq:corr_den} in which we replace $\tilde{P}(\tilde{w}_n, \tilde{w}_{n+\delta n}| L / l(E)) \rightarrow \tilde{P}_{\rm st}(\tilde{w}_n, \tilde{w}_{n+\delta n})$. This demonstrates the saturation of ${\cal C}_n(\delta n)$ at an $E$-independent value for $n_{\rm c, N} \ll \delta n \ll n_{\rm c, A}$.

To find the particular value at which ${\cal C}_n(\delta n)$ saturates, we need to know $\tilde{P}_{\rm st}(\tilde{w}_n, \tilde{w}_{n+\delta n})$. Although the latter function cannot be found analytically, one can straightforwardly obtain it numerically. A possible way to do it, is to first discretize the variables $\tilde{w}_n$ and $\tilde{w}_{n + \delta n}$ in Eq.~\eqref{eq:Pw_nondim}, and then to find an eigenvector with a zero eigenvalue of a matrix corresponding to the differential operator on the right hand side of Eq.~\eqref{eq:Pw_nondim}. Adopting this approach, we obtain ${\cal C}_n(\delta n) \simeq 0.31$, as presented in the main text.

\section{Derivation of the estimate used to find the conductance correlation function ${\cal C}_{n}(\delta n)$ at $\delta n \ll n_{\rm c, N}$}
To find the correlation function ${\cal C}_{n}(\delta n)$ at $\delta n \ll n_{\rm c, N}$ and $\Gamma = 0$, 
we needed to estimate the divergence of the $\vec{w}$\,-particles at two values of electron densities. For this divergence, in the main text we stated $|\vec{w}_{n+\delta n} - \vec{w}_{n}| \sim \sqrt{\delta k_\mu \Delta x}$, where $\Delta x$ is the ``time'' measured with respect to the beginning of the motion cycle, and $\delta k_\mu$ is the variation of the Fermi momentum due to $\delta n$.
Throughout the section, we assume that $\delta n \ll n_{\rm c, N} \ll n_{\rm c, A}$.

We start with the derived in Sec.~\ref{sec:saturation} equations~\eqref{eq:evol_den_1} and \eqref{eq:evol_den_2}. In these equations, $\eta_{{\rm N 0}, n} = \eta_{{\rm N}x}\sin \alpha_{0,n} - \eta_{{\rm N}y}\cos \alpha_{0,n}$ and $\eta_{{\rm N 0}, n + \delta n} = \eta_{{\rm N}x}\sin \alpha_{0,n + \delta n} - \eta_{{\rm N}y}\cos \alpha_{0,n + \delta n}$. The difference between $\alpha_{0,n}$ and $\alpha_{0,n + \delta n}$ stems from the respective difference in the Fermi momentum $k_\mu$ at the two values of density.
Using Eq.~(8) of the main text and recalling that the EC amplitudes change as $\eta_{\rm N}(x) \rightarrow \eta_{\rm N}(x) e^{-2i\delta k_\mu x}$ with the variation $\delta n$, we obtain the following equations for the evolution of $\alpha_{0,n / n + \delta n}$ with $x$:
\begin{subequations}\label{eq:alphas_den_evol}
\begin{align}
    \frac{d\alpha_{0,n}}{dx} &= \vartheta_{\rm R} + \vartheta_{\rm L} - \bigl(\eta_{{\rm N}x} \cos \alpha_{0,n} + \eta_{{\rm N}y} \sin \alpha_{0,n}\bigr),\\
    \frac{d\alpha_{0,n+\delta n}}{dx} &= \vartheta_{\rm R} + \vartheta_{\rm L} + 2\delta k_\mu - \bigl(\eta_{{\rm N}x} \cos \alpha_{0,n + \delta n} + \eta_{{\rm N}y} \sin \alpha_{0,n + \delta n}\bigr).
\end{align}
\end{subequations}
Let us assume that variables $w_n$ and $w_{n+\delta n}$ are on the same trench, and both of them are far away from the trench end-points.
Due to the difference between noises $\eta_{{\rm N0},n}$ and $\eta_{{\rm N0},n + \delta n}$ in Eqs.~\eqref{eq:evol_den_1} and \eqref{eq:evol_den_2}, $w_n$ and $w_{n+\delta n}$ separate in the course of their evolution.
To describe how the separation $\delta w \equiv w_{n + \delta n} - w_{n}$ changes with $x$, it is convenient to derive a Fokker-Planck equation for the distribution function $P(\delta w, \delta \alpha | x)$, where $\delta \alpha \equiv \alpha_{{\rm N0}, n + \delta n} - \alpha_{{\rm N0}, n}$. Application of the standard procedure \cite{vankampen} to Eqs.~\eqref{eq:evol_diff_dens} and \eqref{eq:alphas_den_evol} results in
\begin{equation}
    \frac{\partial P}{\partial x} = \frac{1}{l_0} [1 - \cos (\delta \alpha)] \frac{\partial^2 P}{\partial(\delta w)^2} - 2\delta k_\mu \,\frac{\partial P}{\partial (\delta \alpha)} + \frac{1}{l_0} \frac{\partial^2}{\partial (\delta \alpha)^2}\bigl\{[1 - \cos (\delta \alpha)]P\bigr\}.
\end{equation}
For $\Delta x \gg l_0$ (as measured with respect to the start of a cycle in the $\vec{w}$\,-particle motion), we can solve this equation using an analog of the Born-Oppenheimer approximation, with the slow and fast variables being respectively $\delta \alpha$ and $\delta w$.
The approximate solution reads $P(\delta w, \delta \alpha | x) \approx P_{w}(\delta w | x) P_{\alpha,\rm st}(\delta \alpha)$, where $P_{\alpha,\rm st}(\delta \alpha)$ is the normalized eigenfunction of the operator $-2\delta k_\mu \,\partial_{\delta \alpha} + \frac{1}{l_0}\partial_{\delta \alpha}^2[1 - \cos(\delta \alpha)]$ with a zero eigenvalue. Function $P_w(\delta w|x)$ satisfies
\begin{equation}\label{eq:diff_for_sep}
    \frac{\partial P_w}{\partial x} = \frac{1}{l_0}\langle 1-\cos(\delta \alpha)\rangle_\alpha\frac{\partial^2 P_w}{\partial(\delta w)^2},\quad\text{where}\quad \langle 1-\cos(\delta \alpha)\rangle_\alpha \equiv \int \frac{d(\delta \alpha)}{2\pi} [1-\cos(\delta \alpha)] P_{\alpha,\rm st}(\delta \alpha).
\end{equation}
This equation shows that the separation $\delta w$ growth with $x$ diffusively, with the diffusion coefficient $\langle 1 - \cos (\delta \alpha)\rangle_\alpha / l_0$. To estimate the diffusion coefficient, we find the angular distribution function $P_{\alpha,\rm st}(\delta \alpha)$. A straightforward calculation gives
\begin{equation}
    P_{\alpha, \rm st}(\delta \alpha) = \frac{\cal N}{1 - \cos (\delta \alpha)} \int_{\delta \alpha}^{2\pi} d\rho \exp\Bigl(-2\delta k_\mu l_0 \Bigl[\cot \frac{\delta \alpha}{2} - \cot \frac{\rho}{2}\Bigr]\Bigr),
\end{equation}
where ${\cal N}$ is the normalization constant. 
For small density variations $\delta n \ll n_{\rm c, N}$, the parameter in the exponent satisfies $\delta k_\mu l_0 \ll 1$. 
Under this condition, the distribution function $P_{\alpha,\rm st}(\delta \alpha)$ has a narrow peak at $\delta \alpha \sim \delta k_\mu l_0$, with a power-law tail $P_{\alpha,\rm st}(\delta \alpha) \sim \delta k_\mu l_0 / (\delta \alpha)^2$ (we assume $k_\mu l_0 \ll \delta \alpha \lesssim 1$ in the latter estimate). In fact, it is this tail that determines the average in Eq.~\eqref{eq:diff_for_sep}. 
We estimate $\langle 1 - \cos (\delta \alpha ) \rangle_\alpha \sim \delta k_\mu l_0$ and thus find that the diffusion coefficient for the separation $\delta w$ is $\sim \frac{1}{l_0} \delta k_\mu l_0 = \delta k_\mu$. This leads to the estimate $\delta w \sim \sqrt{\delta k_\mu \Delta x}$ used in the main text to find ${\cal C}_n (\delta n)$ at $\delta n \ll n_{\rm c, N}$.